\documentclass{emulateapj}

\usepackage{textcomp}
\usepackage[utf8]{inputenc}
\usepackage[english]{babel}
\usepackage{amsmath,amssymb,amsfonts,amscd}
\usepackage{graphicx} 
\usepackage{colortbl}
\usepackage{gensymb}
\usepackage{multirow}
\usepackage{tikz}
\usetikzlibrary{shapes}

\usepackage{hyperref}
\hypersetup{colorlinks,citecolor=blue,linkcolor=blue}

\newcommand{\BB}{{\bf B}}
\newcommand{\AAA}{{\bf A}}
\def\nab{\mbox{\boldmath $\nabla$}}
\definecolor{light-gray}{gray}{0.85}

\shorttitle{Origin and evolution of magnetic field in PMS stars}
\shortauthors{C. Emeriau-Viard \& A. S. Brun}

\begin{document}

\title{Origin and evolution of magnetic field in PMS stars: influence of rotation and structural changes}

\author{ Constance Emeriau-Viard \& Allan Sacha Brun}
\affil{Laboratoire AIM Paris-Saclay CEA/DSM - CNRS - Universit\'e Paris Diderot, IRFU/DAp CEA Paris-Saclay, 91191 Gif-sur-Yvette Cedex, France\\
}
\email{constance.emeriau@cea.fr,sacha.brun@cea.fr}


\begin{abstract}
During stellar evolution, especially in the PMS, stellar structure and rotation evolve significantly causing major changes in the dynamics and global flows of the star.
We wish to assess the consequences of these changes on stellar dynamo,
internal magnetic field topology and activity level. 
To do so, we have performed a series of 3D HD and MHD simulations with the ASH code. 
We choose five different models characterized by the radius of their radiative zone
following an evolutionary track computed by a 1D stellar evolution code.
These models characterized stellar evolution from 1 Myr to 50 Myr.
By introducing a seed magnetic field in the fully convective model and spreading its evolved state through all four remaining cases, 
we observe systematic variations in the dynamical properties and magnetic field amplitude and topology of the models.
The five MHD simulations develop strong dynamo field that can reach equipartition state between the kinetic and magnetic energy and 
even super-equipartition levels in the faster rotating cases. 
We find that the magnetic field amplitude increases as it evolves toward the ZAMS. 
Moreover the magnetic field topology becomes more complex, with a decreasing axisymmetric component and a non-axisymmetric one becoming predominant.
The dipolar components decrease as the rotation rate and the size of the radiative core increase.
The magnetic fields possess a mixed poloidal-toroidal topology with no obvious dominant component.
Moreover the relaxation of the vestige dynamo magnetic field within the radiative core is found to satisfy MHD stability criteria.
Hence it does not experience a global reconfiguration but slowly relaxes by retaining its mixed stable poloidal-toroidal topology.
\end{abstract}





\section{Stellar evolution and magnetism}

Stellar rotation is known to significantly change over the pre main sequence (PMS) through angular momentum conservation as young stars contract under the action of gravitation. 
At the very beginning of the PMS, stellar rotation remains constant since the star is in a disk-locking phase until about 3 to 10 Myr, when it decouples from the vanishing disk. 
Then as the star contracts under the influence of gravitation, stellar rotation increases as a consequence of angular momentum conservation until it reaches the zero age main sequence (ZAMS). 
Later, on the main sequence (MS), stellar rotation decreases as contraction stops and magnetic winds start braking the star. 
This is not the only drastic evolution that young stars experience during this phase of stellar evolution as their luminosity also varies by a large factor. 
The internal structure is too strongly impacted as the star evolves along the PMS. 
Indeed starting from a fully convective state, their radiative zone grows continuously due to the ignition of thermonuclear reactions in their deep core, 
such as occupying most of their interior upon their arrival on the ZAMS. 
These major changes impact the star's properties, especially their internal rotation and magnetic field. 

Stellar rotation rate, internal rotation and magnetic field are strongly linked through complex physical processes. 
At the very beginning of the PMS, stars are meant to rotate quite fast since they contract and accrete angular momentum from the disk. 
However observations led by \cite{Bouvier1986,Bouvier2014} show that they only rotate at one tenth of their break-up velocity. 
Hence some physical process prevent stars from spinning up at the very beginning of their PMS evolution. 
Magnetic field is an likely candidate to explain this phenomena as it controls the interaction between the star and its disk. 
Even after the disk-locking phase, magnetic field has a strong link with rotation through wind-braking and core-envelop coupling. 
Magnetic field can also possibly modify the transport of angular momentum in stellar interiors through Maxwell stresses. 
For instance, it has been invoked to explain the flat rotation profile in the radiative interior of the Sun, 
along with other processes such as internal waves \citep[][]{Charbonnel2005,Eggenberger2005}.
It is quite clear that a feedback loop between rotation and magnetic field must exist. 
On one hand the rotation impacts the magnetic field through dynamo process, 
and especially through the shearing of magnetic field lines by the differential rotation of the convective envelop (e.g. the $\Omega$-effect). 
On the other hand, magnetic field topology and amplitude impact braking by the wind.
Evidence of such an influence was studied for instance by \citep[][]{Pizzolato2003,Wright2011,Gondoin2012,Reiners2014,Matt2015,Blackman2016}. 
Such analysis showed a correlation between coronal X-ray emission and stellar rotation in late-type main-sequence stars, revealing the existence of two regimes. 
In the first one, at Rossby number greater than 0.1-0.3, the X-ray emission is well correlated with the rotation period whereas in the second one, at low Rossby number, a constant saturated X-ray to bolometric luminosity ratio is attained. 
This implies that either the surface field or the stellar dynamo, or both, saturates at fast rotation rates, i.e. at low Rossby numbers. 

Stellar magnetic field and internal rotation can also be influenced by internal structure changes.
The correlation between the existence of the radiative core and X-ray emission was studied by \cite{Rebull2006}. 
The results showed that stars with a radiative core have $L_{\rm{X}}/L_{\rm{bol}}$ values that are systematically lower by a factor of 10 than those found for fully convective stars of similar mass.
The flux reduction from fully convective stars to stars with a radiative core is likely related to structural changes that influence the efficiency of magnetic field generation and thus the amplitude and topology of magnetic field.
A correlation between the growth of the radiative core and the reduction of the number of periodically variable T Tauri stars have been established by \cite{Saunders2009}. 
Several surface magnetic maps of accreting T Tauri stars have been published \citep[e.g.][]{Donati2011,Donati2012,Hussain2009}.  
These maps were used by \cite{Gregory2012} to study the influence on stellar magnetic fields of the apparition and growth of a radiative core. 
It has been found that for stars with a massive radiative core, e.g. $M_{core} > 0.4~ M_{star}$, the internal magnetic field is complex, non axisymmetric and has weak dipole components. 
This behavior changes when the radiative core is smaller, e.g. $0 < M_{core} < 0.4~ M_{star}$, the field is less complex and more axisymmetric whereas the dipole component is still weak compared to higher order components. 
As young solar-like stars evolve along the PMS, the internal structure changes from fully convective stars to stars with a radiative core, 
one can expect similarities with low-mass stars behavior, in particular near the M3-M4 transition.
Near the fully convective limit, most of the stars have axisymmetric fields with strong dipole components whereas  
in fully convective stars, the behavior of stellar magnetic fields might even be bistable with a mixture of different geometries and amplitudes \citep[][]{Morin2010}. 
Observations have also shown that solar-like stars possess a magnetic field which is predominantly toroidal for fast rotators \citep[][]{Petit2008,See2016} and that a
 subset may even possess a magnetic cycle. 
To be more precise, some correlations between the period of these cycles and the stellar rotation were advocated during the last decades \citep[][]{Noyes1984,Soon1993,Baliunas1996,SaarBrandenburg1999,Saar2002}. 
These analysis show that $P_{\rm{cyc}}$ increases as $P_{\rm{rot}}$ increases since they found that $P_{\rm{cyc}} \propto P^n_{\rm{rot}}$, where $n$ varies depending on the studies but remains positive. 
However these studies are now reconsidered since they are based on observations of the chromospheric cycle that may differ from the magnetic one \cite{See2016}, as it is the case in the Sun \citep[][]{Shapiro2014}. 
Recent non-linear simulations led by \cite{Strugarek2017} seem to show that the $P_{\rm{cyc}}$ vs $P_{\rm{rot}}$ relation may not be so straightforward.
Indeed in some of dynamo models, the cycle period decreases while the rotation rate increases \citep[see also][]{Jouve2010}.

Since during the stellar evolution along the PMS, the radiative zone of the star increases, 
we also wish to know how the magnetic field evolves in the radiative zone as convective dynamo action does not support it anymore. 
These magnetic fields are observed in massive stars, since their envelop is radiative, where they are often oblique dipoles. 
In solar-like stars, knowledge of magnetic field in radiative core is important even if it is buried under the dynamo field. 
Indeed it is a candidate for the transport of angular momentum in the stellar core that can explain the rotation profile  
observed by helio- and asterioseismology \citep[][]{Schou1998,Garcia2007,Benomar2015}. 
These magnetic fields left by a convective zone in a stably stratified zone are called \textit{fossil fields}. 
Studies led by \cite{Tayler1973}, \cite{MarkeyTayler1973}, \cite{Braithwaite2007} and \cite{Brun2007} showed that purely poloidal or toroidal magnetic fields are unstable in such stably stratified zones. 
\cite{Tayler1980} proposed that the field needs a mixed configuration to be stable in radiative regions. 
This statement was confirmed by numerical simulations and theoretical works \citep[e.g.][]{BraithwaiteSpruit2004,Duez2010}. 
\cite{Braithwaite2008} introduced a constraint on the relative amplitude of poloidal and toroidal field in a stable fossil field: $E_{\rm{pol}}/E_{\rm{tot}} < 0.8$. 
Hence it is interesting to assess if the left over magnetic field is  stable or if it must relax to a different configuration. 

The origin of stellar magnetic activity and regular cycle is supposed to be linked to a global scale dynamo acting in and at the bottom of the convective envelop \cite[][]{Parker1993}. 
This dynamo is a complex dynamical process that can be understand using the \textit{mean field} theory \cite[][]{Moffatt1978}. 
Its main ingredients are the $\Omega$-effect, and the helical nature of small scales convective motions, called the $\alpha$-effect
\citep[e.g.][]{Parker1955,Parker1977,SteenbeckKrause1969}. 
Alternative mechanisms based on the influence of surface magnetic fields have also been developed, as in the Babcock-Leighton dynamo process \citep[][]{Babcock1961,Leighton1969,Choudhuri1995,Charbonneau2005,JouveBrun2007,MieschBrown2012}. 
These theories allow us to reproduce large scales behavior of the magnetic fields in solar-like stars, such as cycles, but lack an explicit treatment of turbulence and many non linear effects.
Thus we cannot rely only on the mean field theory and 3D simulations are an ideal tool to perform such studies.
The earliest non-linear, turbulent and self-consistent works on stellar convection and dynamo models in spherical geometry, first with the Boussinesq approximation then with the anelastic hypothesis, were done during the mid 80's by P. Gilman and G. Glatzmaier \citep[][]{Gilman1983,Glatzmaier1985a,Glatzmaier1985b}. 
During the last three decades, several groups developed global or wedge-like MHD simulations of convective dynamo, including the ASH code \citep[][]{Brun2004,Browning2006,Dobler2006,Browning2008}, the PENCIL code \citep[][]{Warnecke2013,Kapyla2013,Guerrero2013}, the EULAG code \citep[][]{Ghizaru2010,Racine2011,Charbonneau2013} and the MAGIC code \citep[][]{Christensen2006}.
Observations, theoretical models and 3D numerical simulations enable us to improve our understanding of the MHD processes in solar-like stars. 
Nowadays we believe that magnetic field in convective envelope of solar-like stars is due to a dynamo with two separate ranges of spatial and temporal scales. 
The global dynamo explains regular cycles and butterfly diagrams and might be seated at the base of the convective zone and in the tachocline. 
A local dynamo is likely to be at the origin of the rapidly varying and smaller scale magnetism. 
All these phenomena take place in the convective zone of the star as all the dynamo theories cited above need convection motions and differential rotation as essential ingredients for regenerating magnetic field. 

In our study, we compute 3D global magnetohydrodynamical models of PMS stars at different stages of their early evolution to understand the impact of both structural change and rotation on the internal mean field flows and magnetic field. 
We choose five different models with specific stellar parameters as presented in section \ref{ModelSetup}. 
In section \ref{HDProgenitors}, we present the hydrodynamical progenitors of our five models. 
We study the influence of both stellar rotation rate and internal structure on the internal flows and convective motions. 
Thereafter, we introduce magnetic fields in these HD progenitors. 
Thus, we can observe the resulting changes on the hydrodynamical flows and convection (see section \ref{DynamoAction}). 
In section \ref{ExplainingTheDynamics}, we analyze the amplitude and topology of the magnetic field. 
We also look at the magnetic field generation in the convective zone and how it evolves as the models go through its evolution and as rotation rate and internal structure change.
We finally follow the evolution of the magnetic field in the radiative zone by looking at its stability in the core and its relaxation along the stellar evolution.
In section \ref{Conclusion}, we discuss the results and we conclude. 

\section{Model setup}\label{ModelSetup}

To study the evolution of magnetic field during the PMS, we compute 3D global magnetohydrodynamical models of one solar mass star. 
However stellar evolution in the PMS lasts for several Myrs whereas 3D MHD simulations can only compute stellar evolution for several hundreds of years.
Since we cannot compute our simulations for secular time-scales, we select specific models that represents the important stages of the PMS. 
To characterize these models, we need adequate values for luminosity, rotation rate, radii ... 
These physical values are given by 1D stellar evolution models that were computed with the STAREVOL code \citep[][]{Amard2016}.

\subsection{1D evolution}

\begin{figure}[t]
\begin{center}
\includegraphics[scale=0.5]{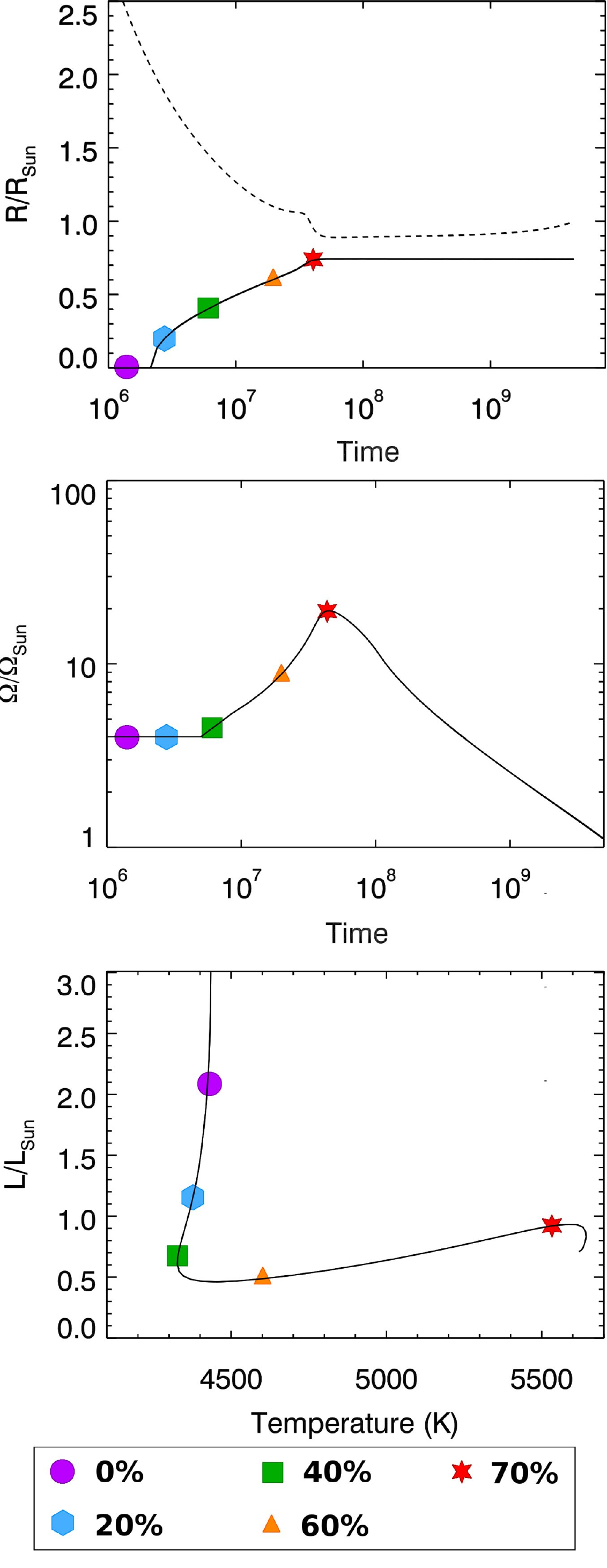}
\caption{\label{1D}Choice of 3D ASH models based on stellar evolution computed with STAREVOL code. 
\textit{Top} : Evolution of stellar radius and radius of the radiative core during the PMS and the MS. 
\textit{Middle} : Stellar rotation rate as a function of time. 
At the beginning of the PMS it is constant since star is still in the disk-locking phase. 
Then it increases as the star contracts under the effect of gravity until the ZAMS. 
Stellar rotation rate starts decreasing as the stellar contraction stops and magnetic wind brakes the star. 
\textit{Bottom} : Relation between luminosity and temperature through stellar evolution. Luminosity decreases, as star ages, for the 4 first models and increases until star arrives at the ZAMS.}
\end{center}
\end{figure}

We study the evolution of a 1 $M_{\odot}$ solar-like star during PMS from a fully convective progenitor to the ZAMS.
This evolution drastically changes the main stellar parameters: radius, size of the radiative core, rotation rate, luminosity and temperature, as represented in Figure \ref{1D}. 
On the upper plot, we can observe that during this evolution the radial structure of the star evolves.
At the very beginning of the PMS, no nuclear reactions occur in the core of the young star.
The energy of PMS stars is due to the release of gravitational potential energy. 
Convection is efficient enough to transport energy in the stellar interior. 
Thus, at the very beginning of the PMS, young stars are fully convective.
Since there is no internal process to counterbalance the gravitational contraction, the radius of the star decreases and we can see in Figure \ref{1D} that 
the stellar radius contracts from 2.5 $R_{\odot}$ to $~$1 $R_{\odot}$.
As the outer radius becomes smaller, temperature and density increase at the center of the star and the opacity of the core drops as : $\kappa \propto \rho T^{-7/2}$. 
When the opacity becomes small enough, the radiative zone appears at the center of the star. 
We see on the Figure that this radiative zone grows up to 70\% of the outer radius as the star reaches the ZAMS and remains stable later on the main sequence. 

As the star contracts, the rotation rate of the star also changes through angular momentum conservation \citep[][]{GalletBouvier2013}. 
This evolution shows three different main phases that can be observed in Figure \ref{1D} (middle panel). 
First of all, we see a locking phase with the protostellar disk where stellar rotation remains constant until around $4~\text{Myr}$.
Then the contraction of the star impacts its rotation rate that increases due to angular momentum conservation until the end of the ZAMS, at $50~\text{Myr}$ in our study. 
In the third and last phase, the star rotation decreases following the Skumanich trend: $\Omega_0 \propto t^{-1/2}$ due to the influence of wind-braking (\cite{Reville2015}).
This modeling of the stellar rotation rate has free parameters such as the initial period at 1Myr, the time coupling between the core and the envelope, the lifetime of the disk and the scaling constant of the wind-braking law. 
Hence, slow, median and fast rotators can be modeled, as seen in \cite{GalletBouvier2013} and \cite{GalletBouvier2015}. 
For our study, we choose an intermediate rotational evolution profile that comes from STAREVOL models \citep[][]{Amard2016}.

1D simulations, such as STAREVOL models, can compute stellar evolution over secular time-scales. 
However, they are restricted in space since they do not consider the angular dependencies. 
On the contrary, 3D global simulations give a more precise understanding of the physical processes that take place in the star, 
but they do not yet enable the study of star's life over several million years. 
To study stellar evolution, we use STAREVOL structures to perform relevant 3D models at key instants in the star's evolution. 
We decided to select five models such that the radiative core radii (in stellar radius unit) are well distributed, almost every 20\%, 
and the ratio between the rotation rate of two consecutive models is smaller than two. 
Such radial structures enable us to create reference states and thus initialize our 3D ASH simulations.
We now turn to describe the 3D setup.

\subsection{3D numerical models}

\subsubsection{Computational methods}

The 3D full sphere simulations of the evolution of solar-type stars during the PMS to the ZAMS presented here are computed with the ASH code \citep[][]{Clune1999,Brun2004,Alvan2014}. 
This code evolves the Lantz-Braginski-Roberts (LBR) form of the anelastic MHD equations for a conductive plasma in a rotating sphere \citep[][]{Jones2011}. 
The anelastic approximation filters fast magnetoacoustic waves but Alfvén and slow magnetoacoustic waves remain. 
In the ASH code, the equations are non-linear in velocity and magnetic field, and are linearized for thermodynamical variables.
These variables are separated into fluctuations $X$ and a reference state $\bar{X}$, which only depends on the radial coordinate and evolves slightly over time. 
We assume the linearized equation of state: 
\begin{equation}
\label{eq::state}
\frac{\rho}{\bar\rho} = \frac{{P}}{\bar{{P}}} -
  \frac{{T}}{\bar{{T}}} = \frac{{P}}{\gamma\bar{{P}}} -\frac{{S}}{c_p}\hbox{,}
\end{equation}
with the ideal gas law: 
\begin{equation}
\label{eq::idealgas}
\bar{P} = \mathcal{R}\bar{\rho} \bar{T},
\end{equation}
where $\rho$, $P$, $T$, $S$ have their usual meaning, $c_p$ is the specific heat per unit of mass at constant pressure, $\gamma$ is the adiabatic exponent and $\mathcal{R}$ is the ideal gas constant.
The continuity equation is: 
\begin{equation}
\label{eq::continuity}
\nab \cdot \left( \bar{\rho} \mathbf{v}\right) = 0
\end{equation}
with $\mathbf{v} = \left(v_r,v_{\theta},v_{\varphi}\right)$ is the local velocity in spherical coordinates and 
$\left(r,\theta,\varphi\right)$ is the spherical frame rotating at a constant velocity $\Omega_0 \hat{\mathbf{e}}_z$.
Under the LBR formulation, the momentum equation can be written as: 
\begin{equation}
\label{eq::momentum}
\begin{matrix}
\displaystyle{\bar{\rho} \left( \frac{\partial \mathbf{v}}{\partial t} + \left(\mathbf{v} \cdot \nab\right) \mathbf{v} \right)} & = & \displaystyle{- \bar{\rho} \nab \tilde{\omega} - \bar{\rho} \frac{S}{c_P} \mathbf{g} - 2 \bar{\rho} \boldsymbol{\Omega}_0 \times \mathbf{v}} \\
& & \displaystyle{- \nab \cdot \mathcal{D} + \frac{1}{4\pi} \left( \nab \times \BB \right)\times \BB}
\end{matrix}
\end{equation}
where $\tilde{\omega} = P/\bar{\rho}$ is the reduced pressure that replaces pressure fluctuations in LBR formulation, $\mathbf{g}$ is the gravitational acceleration, $\BB = \left( B_r, B_{\theta} , B_{\varphi}\right)$ is the magnetic field and $\mathcal{D}$ is the viscous stress tensor given by:
\begin{equation}
\mathcal{D}_{ij} = - 2 \bar{\rho} \nu\left[ e_{ij} - 1/3 \left(\nab \cdot \mathbf{v} \right) \delta_{ij}\right]
\end{equation}
with $e_{ij} = 1/2 \left(\partial_i v_j + \partial_j v_i\right)$ is the strain rate tensor and $\delta_{ij}$ the Kronecker symbol.
Since we study magneto-hydrodynamical simulations, we need to consider the flux conservation equation for the magnetic field:
\begin{equation}
\label{eq::divergence}
\nab \cdot \BB = 0
\end{equation}
 and the induction equation
\begin{equation}
\label{eq::induction}
\frac{\partial \BB}{\partial t} = \nab \times \left( \mathbf{v} \times \BB \right) - \nab \times \left( \eta \nab \times \BB\right)
\end{equation}
with $\eta$ the magnetic diffusivity.  
The magnetic field and mass flux are decomposed into
\begin{equation}
\begin{matrix}
\bar{\rho}\mathbf{v} & = & \displaystyle{\nabla \times \nabla \times \left(W \mathbf{\hat{e}}_r\right) + \nabla \times \left(Z \mathbf{\hat{e}}_r\right)} \\
\BB        & = & \displaystyle{\nabla \times \nabla \times \left(C \mathbf{\hat{e}}_r\right) + \nabla \times \left(A \mathbf{\hat{e}}_r\right)} \\
\end{matrix}
\end{equation}
to ensure that they remain divergenceless to machine precision throughout the simulation. 
Finally the internal energy conservation is: 
\begin{eqnarray}
\label{eq::energy}
\begin{aligned}
\displaystyle{\bar{\rho} \bar{T}} & \displaystyle{\left[\frac{\partial S}{\partial t} + \mathbf{v} \cdot \nab \left( \bar{S} +S\right)\right]} \\
& = \displaystyle{ \bar{\rho} \epsilon + \frac{4\pi\eta}{c^2} \mathbf{j}^2 + 2 \bar{\rho} \nu \left[ e_{ij} e_{ij} - 1/3 \left(\nab \cdot \mathbf{v}\right)^2\right]} \nonumber \\
& + \displaystyle{ \nab \cdot \left[ \kappa_r \bar{\rho} c_p \nab \left( \bar{T} + T \right) + \kappa \bar{\rho} \bar{T} \nab S + \kappa_0 \bar{\rho} \bar{T}  \nab \bar{S}\right]} \nonumber \\
\end{aligned}
\end{eqnarray} 
where $\nu$ and $\kappa$ are effective eddy diffusivities that represent momentum and heat transport by subgrid-scale (SGS) motions, $\kappa_r$ is the radiative diffusivity and $\mathbf{j} = c/4\pi \left(\nab \times \BB\right)$ is the current density.
The diffusivity $\kappa_0$ also represents a subgrid process. 
It is fitted to have the unresolved eddy flux carrying the stellar flux outward the top of the domain. 
This flux drops exponentially with depth since it should play no role inside the radiative zone.
In the energy equation, we have a volume heating term $\bar{\rho} \epsilon$ that represents the energy generation by nuclear fusion in the core of the star. 
We can fit the nuclear reaction rate by a simple model $\varepsilon = \varepsilon_0 \bar{T}^n$.
For each model, we adjust both parameters $\varepsilon_0$ and $n$ such as to have a heating source in agreement with the corresponding 1D STAREVOL model, see Table \ref{star_parameters}.

\begin{table*}[t]
\begin{center}
\caption{Stellar parameters of our numerical simulations}\label{star_parameters}
\vspace{0.2cm}
\begin{tabular}{cccccccc}
\tableline
\tableline
 Case  & Stellar radius & $D$                   & Radiative radius & Luminosity    & Rotation rate      & $\varepsilon_0$        & $n$ \\ [0.5ex]
       & ($R_{\odot}$)  & ($cm$)                & ($R_*$)          & ($L_{\odot}$) & ($\Omega_{\odot}$) &                     &     \\ [0.8ex]
\tableline 
\tableline 
FullConv & $2.44$         & $1.7 \times 10^{11}$  & $0$              & $2.09$        & $3.5$              & $2.7$               & $1$   \\ 
20 \%    & $1.87$         & $1.04 \times 10^{11}$ & $0.2$            & $1.16$        & $3.5$              & $1.3$               & $1$   \\
40 \%    & $1.45$         & $6 \times 10^{10}$    & $0.4$            & $0.66$        & $4.47$             & $0.24$              & $1.7$ \\
60 \%    & $1.10$         & $3.6 \times 10^{10}$  & $0.6$            & $0.49$        & $8.74$             & $9 \times 10^{-3}$  & $3$   \\
70 \%    & $1$            & $2.08 \times 10^{10}$ & $0.7$            & $1$           & $14.0$             & $10^{-12}$          & $9$   \\ [0.5ex]
\hline
\tableline 
\tableline 
\end{tabular}
\end{center}
\textbf{Note:} Seven stellar parameters that characterize the five stars we choose to model (see Figure \ref{1D}) : radius, thickness of the convective envelop ($D = r_{\rm{top}} - r_{\rm{bcz}}$), radius of the radiative core, luminosity, rotation rate and $\varepsilon_0$ and $n$ characterizing the nuclear reaction rate : $\varepsilon = \varepsilon_0 \bar{T}^n$. 
In all our models, we choose to fix the outer radius of the simulation to 96\% of the stellar radius.
The $70\%$ simulation has the same internal structure than the Sun. 
The main difference between this model and the Sun is its rotation rate which is 14 times greater in our model.
\end{table*}

\subsubsection{Problem setup and boundary conditions\label{setup}}

\begin{figure}[t!]
  \begin{center}
    \includegraphics[scale=0.35]{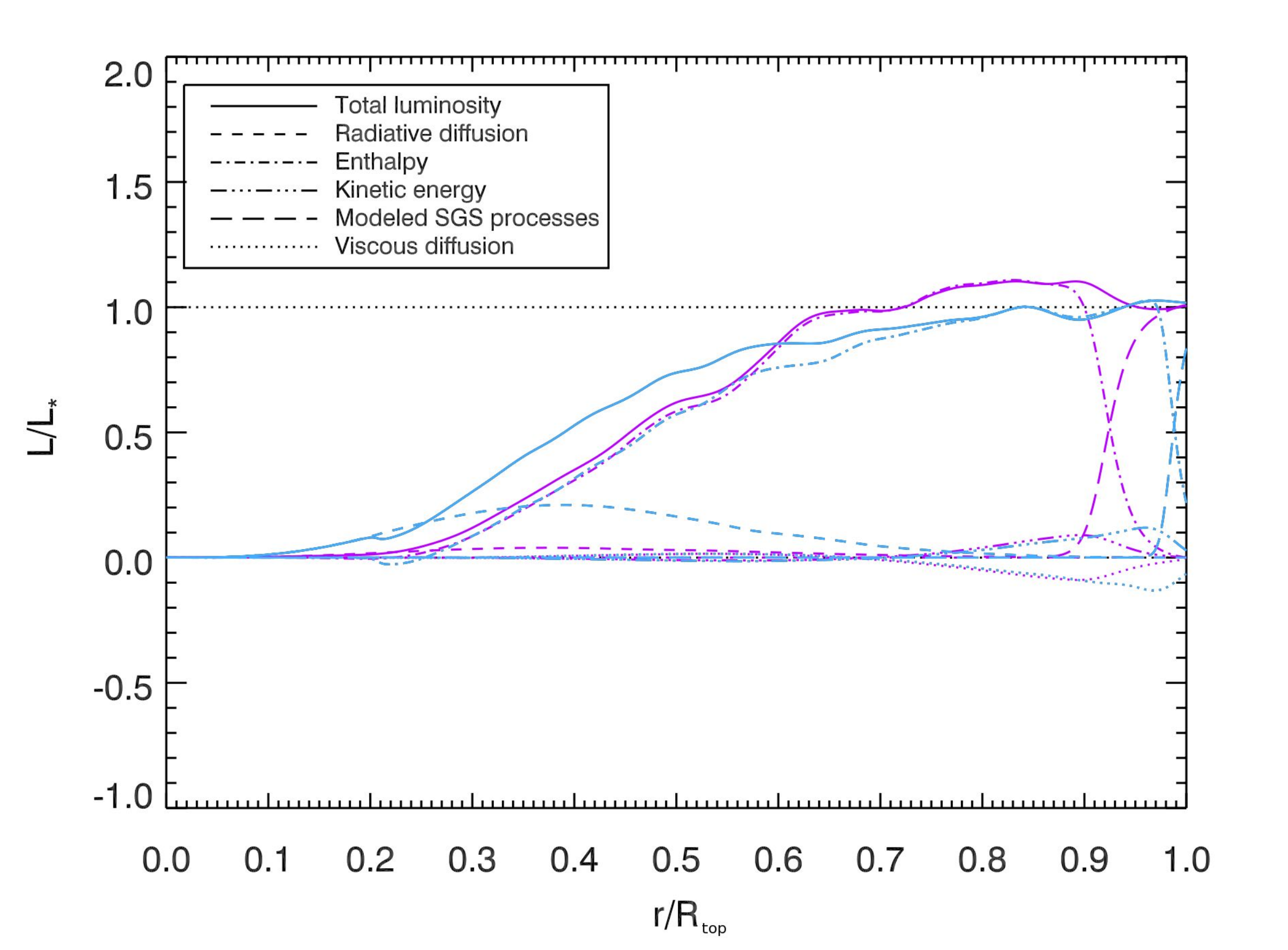}
    \caption{\label{FB}Radial luminosity balance for two models : FC (purple) and 20\% (blue). 
      Luminosities and radii are normalized to their respective stellar values. 
      In these balances, we show the contribution to the total luminosity (solid line) from radiative diffusion (short dashes), enthalpy (dot-dashed), kinetic energy (three-dot-dashed), modeled SGS processes (long dashes) and viscous diffusion (dot) averaged over 400 days.
      We notice that in the center of the star, flux luminosity is greater in the model which have a small radiative zone. 
      Indeed it is sustain by the radiative flux (dashed) that is very low in the fully convective star.
}
  \end{center}
\end{figure}

ASH is a large eddy simulation (LES) code with a SGS treatment for motions whose scales are below the grid resolution of our simulations.
These unresolved scales are modeled with diffusivities $\nu$, $\kappa$ and $\eta$ that represent transport of moment, heat and magnetic field in those small scales.
The eddy thermal diffusivity $\kappa_0$ that drives the mean entropy gradient is computed separately and occupies a tiny region in the upper convection zone (dashed plot in Figure \ref{FB}).
This diffusivity transports heat through the outer surface where radial convective motions vanish. 

The radial structure of velocity, magnetic and thermodynamical variables is computed with a fourth-order finite differences while angular structure is computed with a pseudo-spectral method with spherical harmonics expansion. 
Time evolution is solved by a Crank-Nicolson/Adams-Bashorth second-order technique, advection and Coriolis been computed thanks to Adams-Basforth part and diffusion and buoyancy terms thanks to the semi-implicit Cranck-Nicolson scheme \citep[][]{Clune1999}.

The domain of our simulations goes from the center of the star to $96\%$ of the stellar radius for each case considered in this study. 
Indeed ASH code does not compute 3D simulations up to 100\% of the stellar surface since more complex equation of state and very small convection scales would required extreme resolution. 
Moreover, except for the first one, all our models have two zones with a convective envelope and a radiative core. 
Since our models are full-sphere, boundary conditions only have to be imposed at the surface of the star and regularization of the solution is done at the center of the star, as described in \cite{Alvan2014}. 
The velocity boundary conditions are impenetrable and torque-free: 
\begin{equation}
\label{eq::vel_boundary}
\left \{
\begin{matrix}
\left.v_r \right|_{r_{top}} & = & 0 \\
\\
\displaystyle{\left.\frac{\partial}{\partial r} \left( \frac{v_{\theta}}{r}\right)\right|_{r_{top}}} &  =  & \displaystyle{ \left.\frac{\partial}{\partial r} \left( \frac{v_{\varphi}}{r}\right)\right|_{r_{top}}} & = & 0 \\
\end{matrix}
\right . .
\end{equation}
We also fix a constant heat flux
\begin{equation}
\label{eq::S_boundary}
\left. \frac{\partial S}{\partial r} \right|_{r_{top}} = 0 \quad\hbox{and}\quad  \left. \frac{\partial \bar{S}}{\partial r} \right|_{r_{top}} = cste .
\end{equation} 
Finally, we want the surface magnetic field $\BB$ to match an external potential field $\Phi$ that implies:
\begin{equation}
\label{eq::B_boundary}
\BB = \nab \Phi \quad \hbox{ and }\quad \nab^2 \Phi|_{\rm{surface}} = 0.
\end{equation}

\begin{figure}[t]
  \hspace*{0.45cm}
  \includegraphics[scale=0.4]{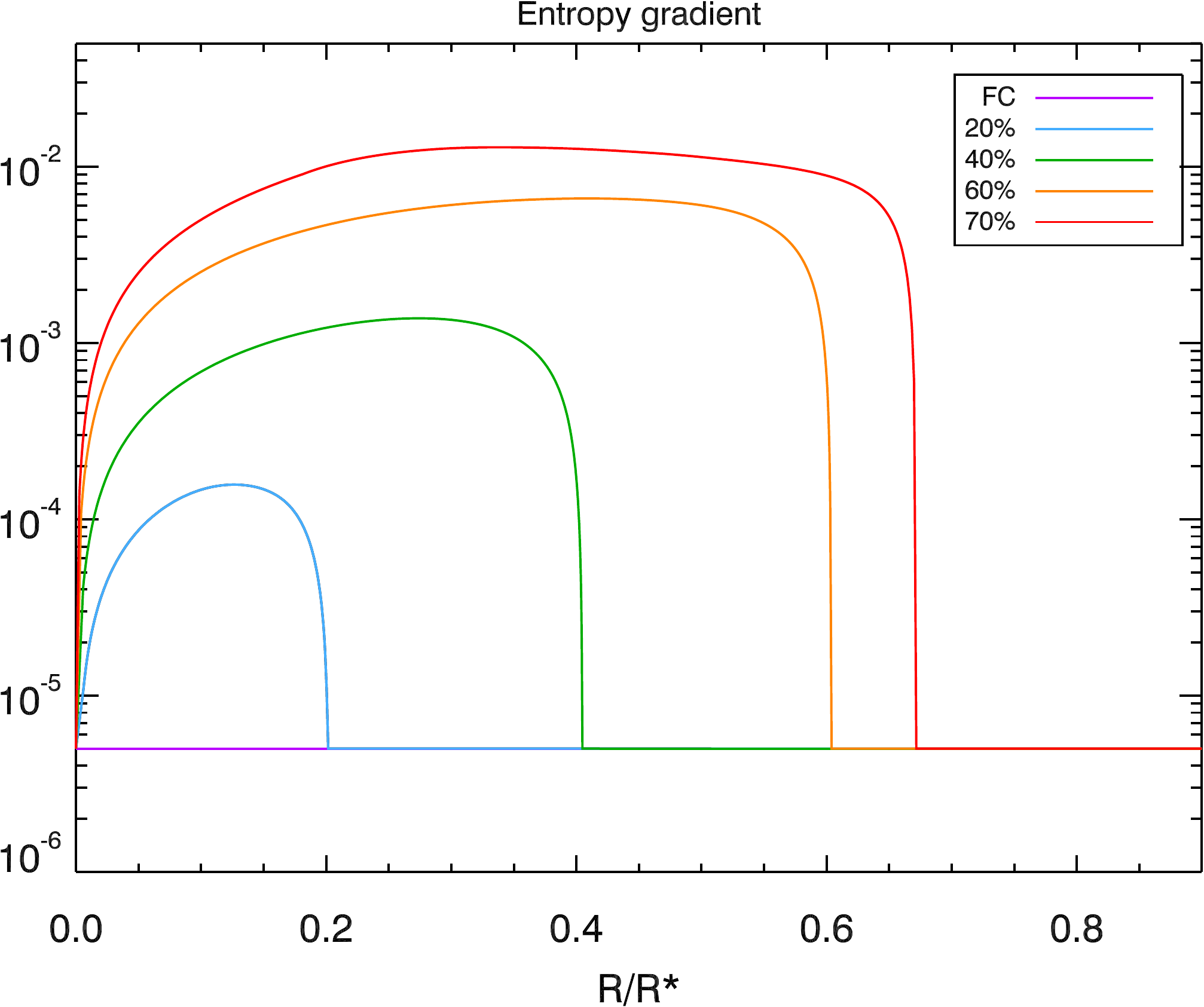}
  \includegraphics[scale=0.4]{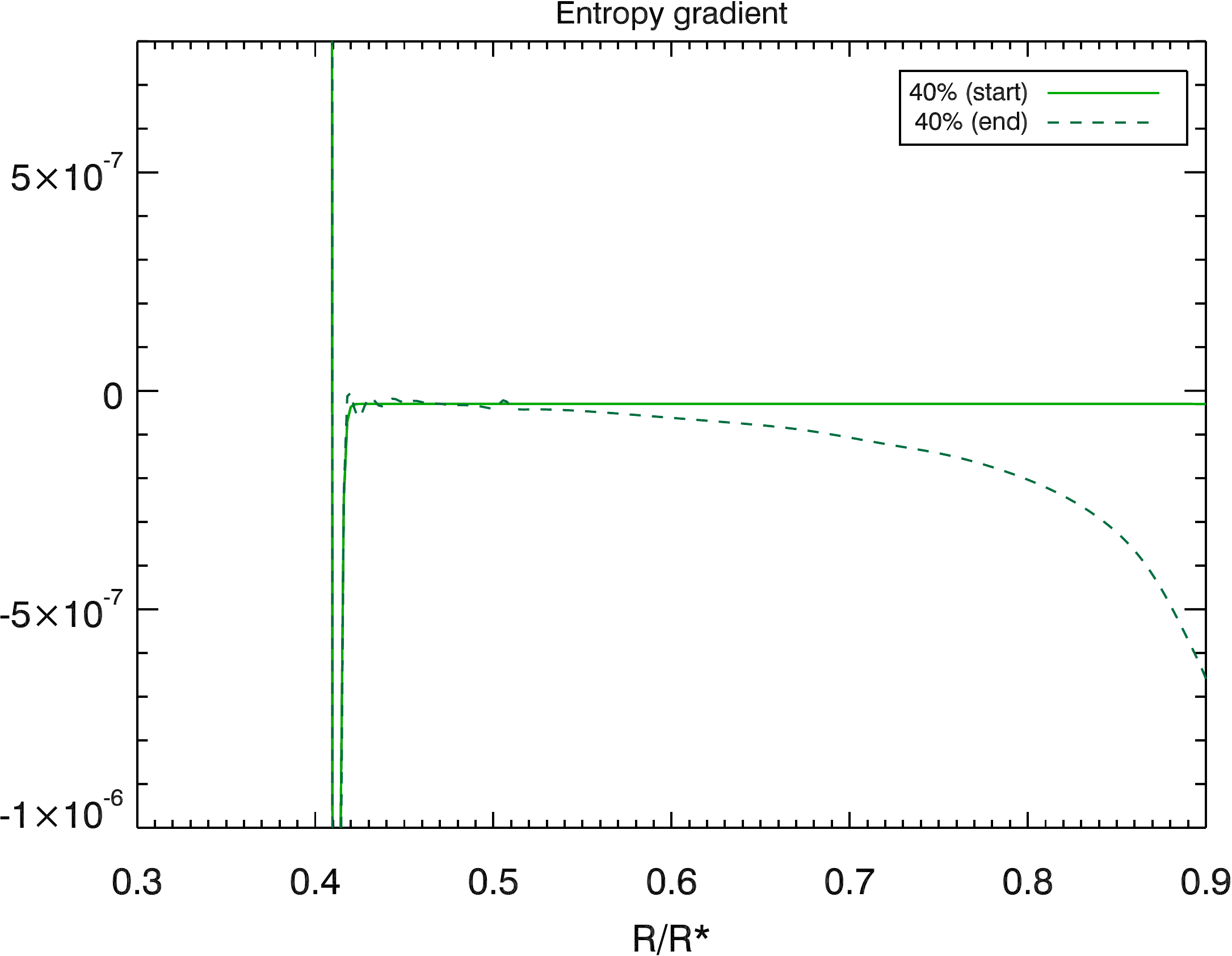}
  \caption{\label{entropy_gradient}Zoom, on the radiative core and the tachocline, of the radial evolution of the entropy gradient for the five hydrodynamical simulations. 
\textit{Top :} In the radiative core, the entropy gradient is positive. As the radiative core grows in the different models, it also becomes more stratified since the entropy gradient amplitude becomes larger.
\textit{Bottom :} In the convective envelop, the entropy gradient is constant and fixed to a small negative value at initial time, 
then it evolves to become more superadiabatic.
}
\end{figure}

From 1D stellar structure computed by STAREVOL, a 1D Lagrangian hydrodynamical stellar evolution code \citep[][]{Siess2000}, we initialize the 3D ASH simulations. 
The gravitational acceleration is fitted from the STAREVOL models. 
The entropy gradient gives us the internal structure of the star: convection occurs where $d\bar{S}/dr$ is negative and for $d\bar{S}/dr$ positive, we have the radiative zone (see Figure \ref{entropy_gradient}). 
In our models we impose a small constant negative entropy gradient in the convection zone. 
The entropy gradient in the radiative zone is deduced from 1D structure as shown in the upper panel of the Figure \ref{entropy_gradient}. 
In this Figure, we see that as the star evolves along the evolutionary path, the radiative zone grows. 
Moreover it becomes more and more stratified, i.e. the values of entropy gradient are greater. 
We run the simulations over hundreds of convection overturning times and we obtain the flux balance given by the Figure \ref{FB}. 
The luminosity flux can be decomposed in several contributions : radiative diffusion, enthalpy, kinetic energy, modeled SGS processes and viscous diffusion. 
By looking on the blue plots, corresponding to the 20\% model, we notice that in the radiative core, the radiative flux is the main contributor to the total flux and the enthalpy flux is negligible 
whereas in the convective zone, we see that the radiative flux decreases with radius and the convective flux dominates. 
In the fully convective model (purple lines), the radiative flux is very low and the luminosity is mainly sustain by the enthalpy flux. 
In both models, the entropy flux is confined to the surface layer and represent the flux carried by the unresolved motions.

\begin{figure}[ht!]
  \begin{center}
    \includegraphics[scale=0.4]{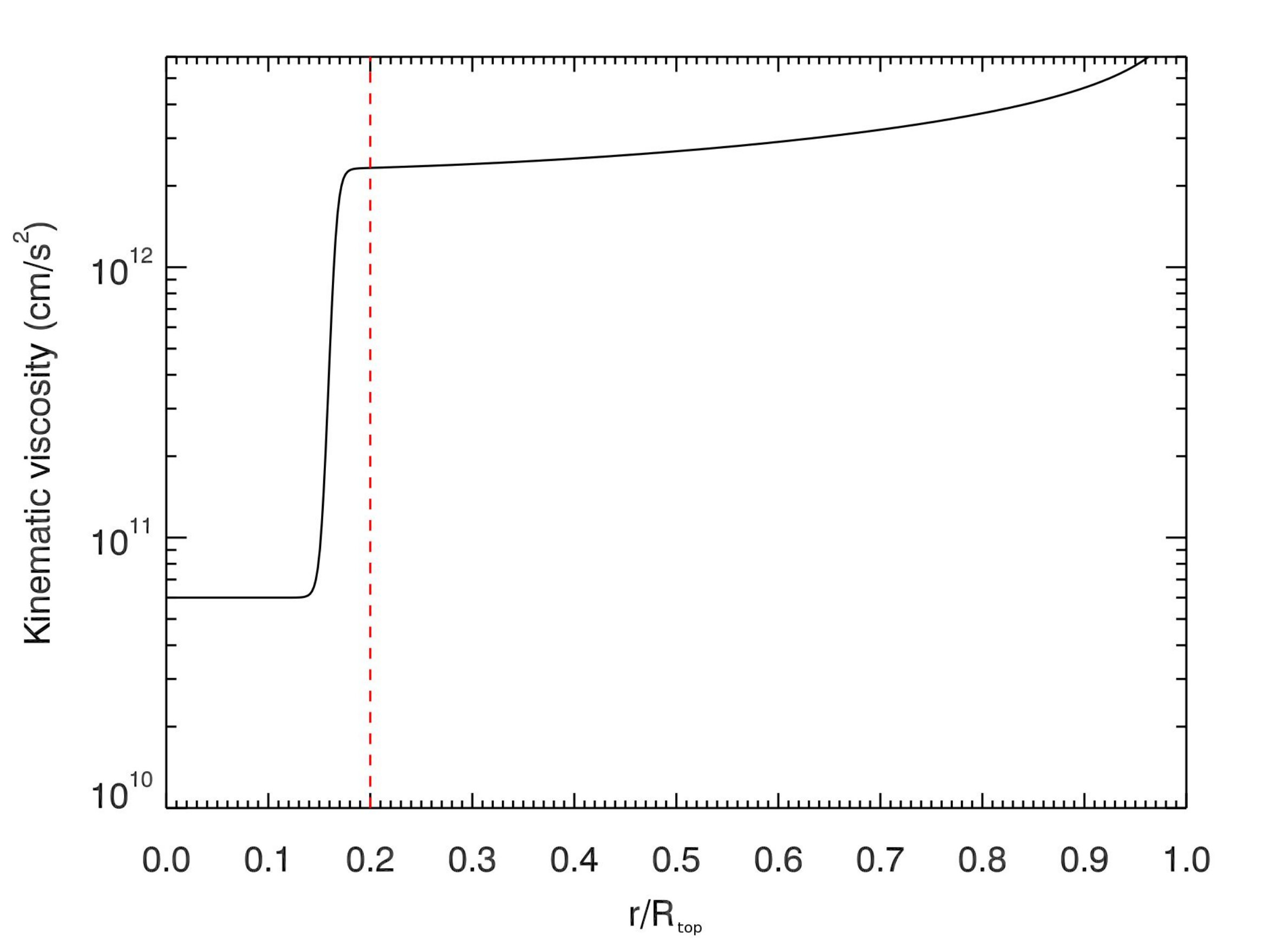}
    \caption{\label{diff}Kinematic viscosity for the 20\% model. 
In the convective envelop, the diffusivity decreases as the depth increases.
The interface between the convective and the radiative zone is characterized by a jump of two order of magnitude of the diffusivity.
In the radiative core, we keep a constant diffusivity.}
  \end{center}
\end{figure}
\begin{table}[h!]
\begin{center}
\caption{Characteristics of diffusivity profiles}\label{t::diff}
\vspace{0.2cm}
\begin{tabular}{cccc}
\tableline
\tableline
Case     & $r_{\rm{t}}$ (cm)     & $\nu_{\rm{top}}$ (cm$^2$.s$^{-1}$) & $\nu_{\rm{rz}}$ (cm$^2$.s$^{-1}$) \\ [0.5ex]
\tableline
\tableline
20 \%    & $2.15 \times 10^{10}$ & $6 \times 10^{12}$                 & $6 \times 10^{10}$                \\
40 \%    & $2.80 \times 10^{10}$ & $2.8 \times 10^{12}$               & $2.8 \times 10^{10}$              \\
60 \%    & $4 \times 10^{10}$    & $1.6 \times 10^{12}$               & $1.6 \times 10^{10}$              \\
70 \%    & $4.4 \times 10^{10}$  & $4 \times 10^{12}$                 & $4 \times 10^{10}$                \\[0.5ex]
\tableline 
\tableline 
\end{tabular}
\end{center}
\textbf{Note:} Quantities caracterising diffusivity profiles.
Radii $r_{\rm{t}}$ locate the jump of diffusivity for the radiative core. 
This jump prevents the spreading of the convective envelope into the radiative core as we compute the simulations.
$\nu_{\rm{top}}$, $\nu_{\rm{rbcz}}$ and $\nu_{\rm{rz}}$ respectively represent the values of the diffusivity at the surface, at the base of the convective envelope and in the radiative core.
\end{table}

For each model, we use the same numerical resolution $N_r \times N_{\theta} \times N_{\varphi} = 500 \times 768 \times 1536$ that gives a horizontal resolution with a maximum spherical harmonic degree $l_{max} = 512$.
We adapt the radial dependence of diffusivities to accommodate to the coexistence of turbulent convective envelope 
with stably stratified radiative interior as the star evolves. 
We use the following formula: 
\begin{equation}
\begin{matrix}
\nu & = & \displaystyle{\nu_{\rm{top}} \left[ c_1 + \left(1-c_1\right) f(r) \left( \frac{\rho}{\rho_0}\right) ^{m}\right]} \\
\end{matrix},
\end{equation}
with $\kappa$ and $\eta$ being calculated with the same type of formula. 
$f$ is a step function:
\begin{equation}
f(r) = \frac{1}{2} \left[\text{tanh}\left(\frac{r-r_{\rm{t}}}{\sigma_{\rm{stiff}}}\right) +1\right],
\end{equation}
with the stiffness $\sigma_{\rm{stiff}} = 0.09 \times 10^{10}\text{ cm}$, the density dependency $m = -0.2$ and $c_1 = 0.01$ remain the same for all the models.
This profile of diffusivity is illustrated in Figure \ref{diff} for the 20\% model. 
Diffusivity is constant in the radiative core, the interface between the two zones is characterized by a jump of two order of magnitude and the diffusivity slightly decreases into the convective envelop. 
The radii $r_{\rm{t}}$, that locates the jump in diffusivities, and $\nu_{\rm{top}}$, that gives the value of the kinematic viscosity at the surface, are referenced in Table \ref{t::diff}. 
We keep a constant Prandlt number, $P_r = \nu/\kappa = 1$. 
In the fully convective star, we choose a magnetic Prandlt number at 0.8, which allows us to get dynamo action. 
When computing the following model, with a small radiative core, we first keep this value of $P_m$ to choose the magnetic diffusivity. 
However the corresponding $R_m$ was not sufficient enough to trigger dynamo action. 
Hence we decrease $\eta$ to reach a sufficient $R_m$ which gives us $P_m = 2$. 
Since in the last three models this value enables us to have dynamo action in the convective envelop, we choose to keep it and to work at $P_m$ constant for all the models with a radiative core. 

\begin{table*}
\begin{center}
\caption{Characteristic numbers}\label{t::numbers} 
\vspace{0.2cm}
\begin{tabular}{ccccccccc}
\tableline
\tableline
Case     & $R_o$                &  $R_a$               &  $T_a$              & $R_e$ & $P_r$ & $R_{e,m}$ & $P_{r,m}$ & $\Lambda$              \\[0.5ex]
\tableline
\tableline
FullConv & $4.4 \times 10^{-3}$ &  $5.9 \times 10^{7}$ & $1.9 \times 10^{8}$ &  61.2 &   1   &   48.96   &     0.8   & $3.2 \times 10^{-3}$   \\
20 \%    & $5.1 \times 10^{-3}$ &  $4.0 \times 10^{7}$ & $1.2 \times 10^{8}$ &  55.6 &   1   &   111.2   &     2     & $9.1 \times 10^{-3}$   \\
40 \%    & $6.4 \times 10^{-3}$ &  $6.0 \times 10^{7}$ & $4.5 \times 10^{7}$ &  43.0 &   1   &    86.7   &     2     & $1.2 \times 10^{-2}$   \\
60 \%    & $6.6 \times 10^{-3}$ &  $2.3 \times 10^{8}$ & $2.9 \times 10^{7}$ &  35.5 &   1   &    71.0   &     2     & $3.6 \times 10^{-2}$   \\
70 \%    & $1.7 \times 10^{-2}$ &  $2.0 \times 10^{8}$ & $4.1 \times 10^{6}$ &  34.8 &   1   &    69.6   &     2     & $3.5 \times 10^{-2}$   \\[0.5ex]
\tableline
\tableline
\end{tabular}
\end{center}
\textbf{Note:} Characteristic numbers for the MHD simulations. 
The characteristic numbers are evaluated at mid-depth of the convective zone. 
These numbers are defined as the Rossby number $R_o = \text{v}_{\rm{rms}}/(2 \Omega_0 D)$, the Rayleigh number $R_a = (-\partial \rho/\partial S) (d\bar{S}/dr)g D^4 / \nu^2$, the Taylor Number $T_a = 4 \Omega_0^2 D^4/\nu^2$, the Reynolds number $R_e = \text{v}_{\rm{rms}} D/\nu$, the Prandtl number $P_r = \nu/\kappa$, the magnetic Reynolds number $R_{e,m} = \text{v}_{\rm{rms}} D/\eta$, the magnetic Prandtl number $P_{r,m} = \nu/\eta$ and the Elsasser number $\Lambda = B_{\rm{rms}}^2/(8\pi\rho \Omega_0 \text{v}_{\rm{rms}} D)$. 
\end{table*}

\begin{figure*}[ht!]
  \begin{center}
    \includegraphics[scale=0.8]{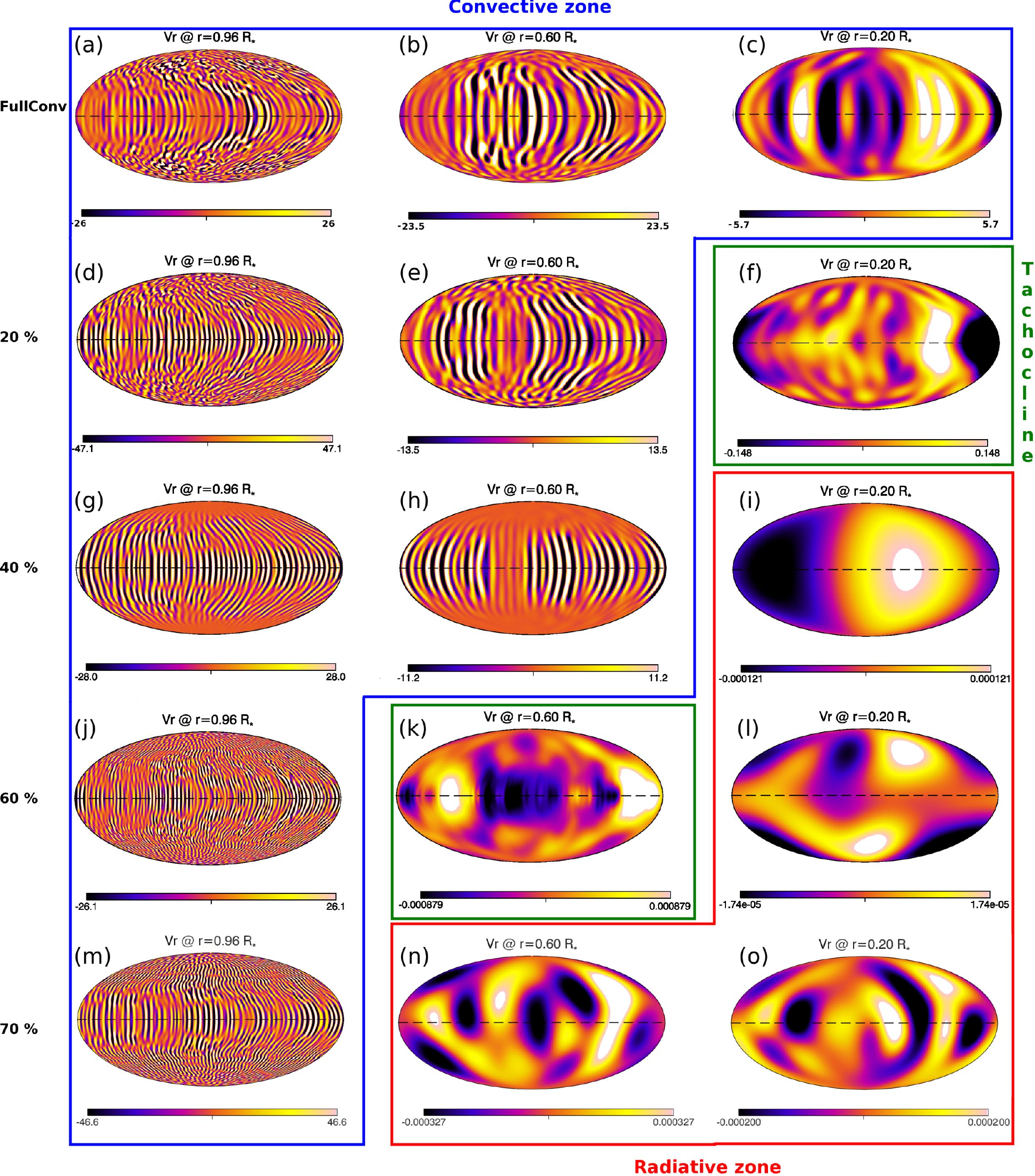}
    \caption{\label{SHSL}Radial velocity for the five HD simulations (rows) at three different depths (columns) : $96\%$, $60\%$ and $20\%$ of stellar radii. Up flows are shown in red/white and down flows in blue/black. 
      At the same depth, the internal structure is not the same in the different models since radiative cores have different sizes. 
      This change, linked with the evolution of stellar rotation rate, leads to important difference for the convection amplitude and patterns.}
  \end{center}
\end{figure*}

Table \ref{t::numbers} shows us the characteristic numbers for our five models. 
The Rayleigh numbers in all models are greater than the critical value which is about $10^4$ for the values of Taylor numbers used her \citep[see][]{Jones2009}. 
The Rossby numbers are significantly smaller than 1. 
Hence, according to \cite{Brun2017}, we can expect the differential rotation profile of our simulations to be solar-like stars with fast equator and slow poles. 
The Elsasser number is smaller than 1 in all models. 
As the star ages, this number increases with the Lorentz begin to counterbalance the Coriolis ones. 

\section{HD progenitors}\label{HDProgenitors}

As seen in Figure \ref{1D}, we choose five models to represent the evolution of one solar mass star during the PMS. 
The different parameters of these models are listed in Table \ref{star_parameters}. 
We then performed ASH 3D simulations of such model stars for which typically 600kh cpus are needed.

\subsection{Internal flow fields}\label{sec::HD}

First of all, we want to analyze the impact of the evolution of internal structures and rotation rates in our five hydrodynamical simulations on the convection patterns and internal flow fields. 
Convection patterns of our five HD simulations evolve as the radiative core grows and the rotation rate increases.

\begin{figure}
  \begin{center}
    \includegraphics[scale=0.4]{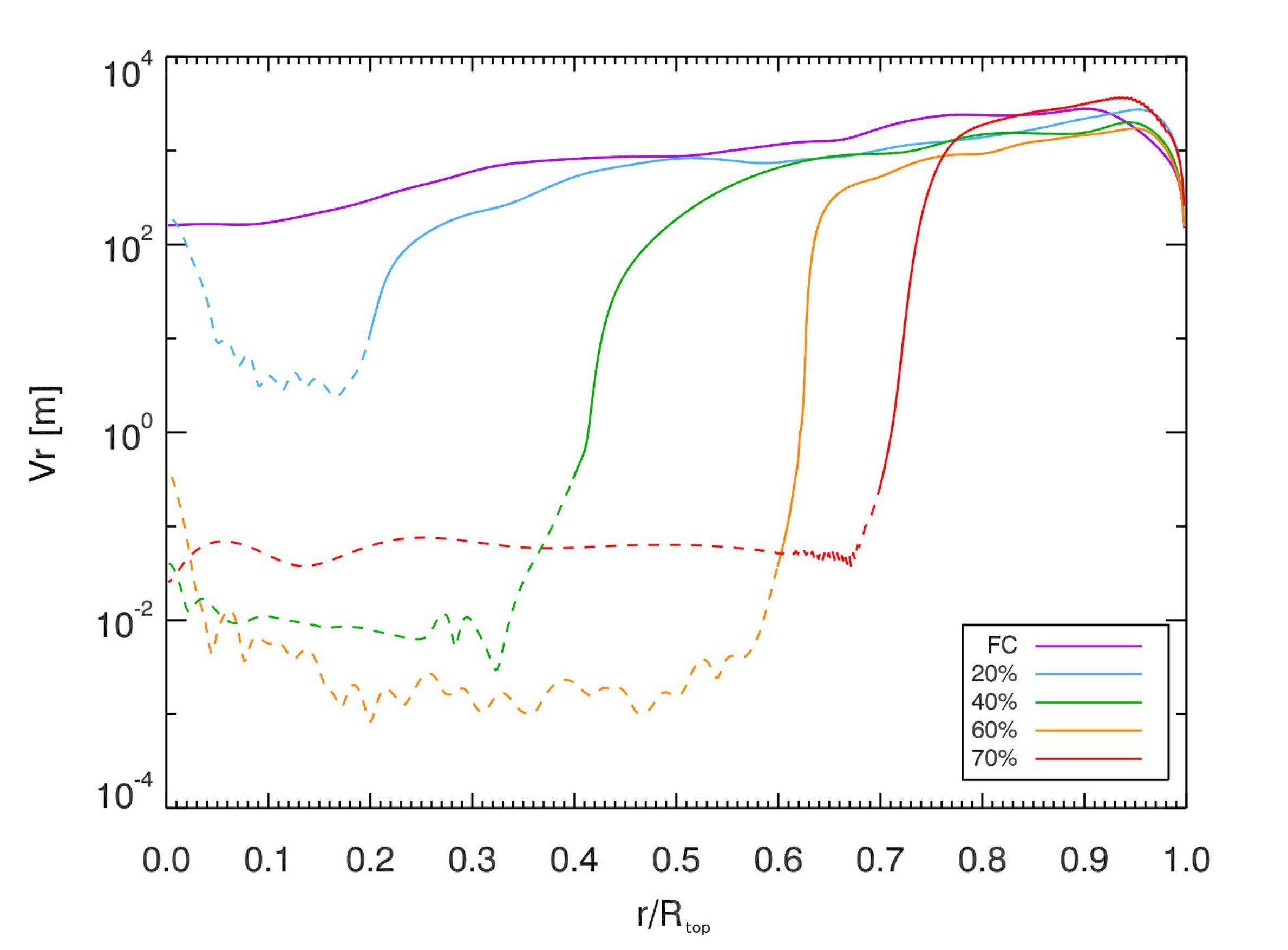}
    \caption{\label{vrms}Radial velocity for the five HD simulations. 
      At the base of the convective envelop, we see a jump in amplitude with values of velocity in the radiative zone that are 100 times, for the 20\% model, to 100 000 times smaller than the value observed in the convective zone.
    }
  \end{center}
\end{figure}
\begin{figure}
  \begin{center}
    \includegraphics[scale=0.7]{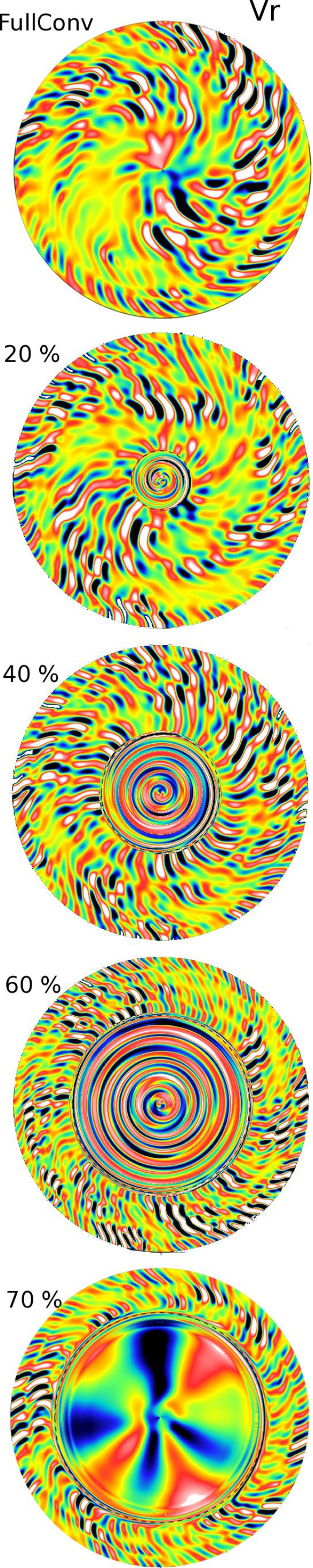}
    \caption{\label{EQSL}Equatorial slices of locally normalized radial velocity, e.g. at each depth the velocity is normalized by the horizontally averaged rms velocity. 
      Downflows are shown in blue/black and upflows are in red/white. 
      The radiative core is delimited by a dotted black line. 
      In that radiative core, we can see the gravito-inertial waves that are excited by the downflows of the convective zone hitting the tachocline. }
  \end{center}
\end{figure}
In Figure \ref{SHSL}, shell slices of radial velocity are shown at three different depths : $96\%$, $60\%$ and $20\%$ of the stellar radius. 
At the surface of our models (\textit{(a)},\textit{(d)},\textit{(g)},\textit{(j)} and \textit{(m)}), we observe two types of convective patterns in all simulations with, at low latitudes, elongated flows that are aligned with the rotation axis, the so-called \textit{banana cells}, and smaller scales at high latitudes. 
These banana cells produce correlations in the velocity field that increase the Reynolds stresses \citep[][]{Gastine2014,Brun2017}. 
The consequence is an acceleration of the equator and a slowdown of the poles that explain the differential rotation profiles of our simulations. 
Convective scales at the surface become smaller as the star evolves along the PMS (from \textit{(a)} to \textit{(m)}). 
These changes in the horizontal direction are due to the increase of the most instable modes $m$ in the convective zone with the rotation rate $\Omega_0$ \citep[][]{Jones2009}. 
After the surface, we look at a deeper shell in the star, i.e. at 60\% of stellar radius. 
In the first three models, \textit{(b)}, \textit{(e)} and \textit{(h)}, we are looking in the convective envelop, since the radiative zone is smaller than $60\%$ of their stellar radius. 
In these simulations, we see that the size of the convective patterns slightly narrows as the star evolves. 
As at the surface, we see banana cells near the equator. 
Moreover these patterns are modulated in amplitude. 
This phenomenon is called \textit{active nest} and is linked to small Rossby and Prandtl numbers \cite[see][]{Ballot2007,Brown2008}. 
The amplitude of convective patterns at that depth slightly decreases compared to the one at the surface. 
As the radiative zone reaches $60\%$ of the stellar radius (plot \textit{(k)}), observations at this depth are not longer focused on the convective envelope but on the tachocline of the star. 
Thus we observe that convection patterns drastically change. 
Small structures disappear but there is still some persistence of the convective patterns, especially at the equator. 
Moreover the amplitude of the convection becomes almost four order of magnitude smaller than the one of the previous models (see Figure \ref{vrms}). 
In the last model \textit{(n)}, we look at a shell in the radiative core of the star and we can see that the last traces of the convective patterns vanish and we are only left with large-scale structures. 
At the lowest depth, near $20\%$ of the stellar radius, convection patterns are completely different depending on the model considered. 
In the fully convective model \textit{(c)}, banana cells can be seen. 
Their horizontal extents are larger than the ones at the surface and at 60\%. 
As the flows go deeper into the stars, they merge and form larger structures. 
The amplitude of the velocity keep the same order of magnitude than in the other depths even if it decreases slightly. 
In the first model with a radiative zone, the observed shell slice \textit{(f)} corresponds to the tachocline of the star. 
As in the mid-depth cut for the 60\% case \textit{(k)}, small convective structures vanish and a large-scale structure emerge. 
The amplitude of convective patterns is smaller than the one observed at the same depth in the convective zone of the fully convective model.
In the last three simulations (\textit{(i)}, \textit{(l)} and \textit{(o)}), the observed shell slice is inside the radiative zone. 
A large-scale structure is predominant in all these models. 
The radial velocity amplitude is much smaller than the one observed in the first two models.

Radial velocity gives a good picture of the convection flows and patterns that occur in our 3D stellar simulations.
In Figure \ref{EQSL}, we show an equatorial slice of the radial velocity for each hydrodynamical model. 
In all the models that have a radiative zone, gravito-inertial waves can be observed in that core. 
These waves are generated by the downflows of the convective envelop. 
These flows come from the surface to the base of the convective zone, where they hit the radiative zone and excite waves in there \citep[][for a detailed discussion of such phenomena in 3D]{Brun2011,Alvan2014,Alvan2015}.
In the convective envelop, changes in convective patterns are observed in the different models. 
They are due to the changes in rotation rate \citep[][]{Jones2009} and in aspect ratio size of the convective envelope. 
At first, we focus on the differences between the first two models since they have the same stellar rotation rate : the main difference between the models is the size and geometry of the convective envelop. 
By comparing these models, we see that the size of convective patterns do not change much through the change of internal structure and the appearance of a radiative core of 20\%. 
In the following models, rotation rate increases as the star contracts.
We notice that both radial and horizontal extents of the convective patterns decrease as the rotation rate increases. 
Horizontal changes are coherent with the ones observed on the shell slices and with the increase of rotation rate discussed by commenting Figure \ref{SHSL}.
Changes on the radial direction may be linked to the changes in geometry and size on the convection zone. 

\begin{figure}[t]
  \begin{center}
    \includegraphics[scale=0.65]{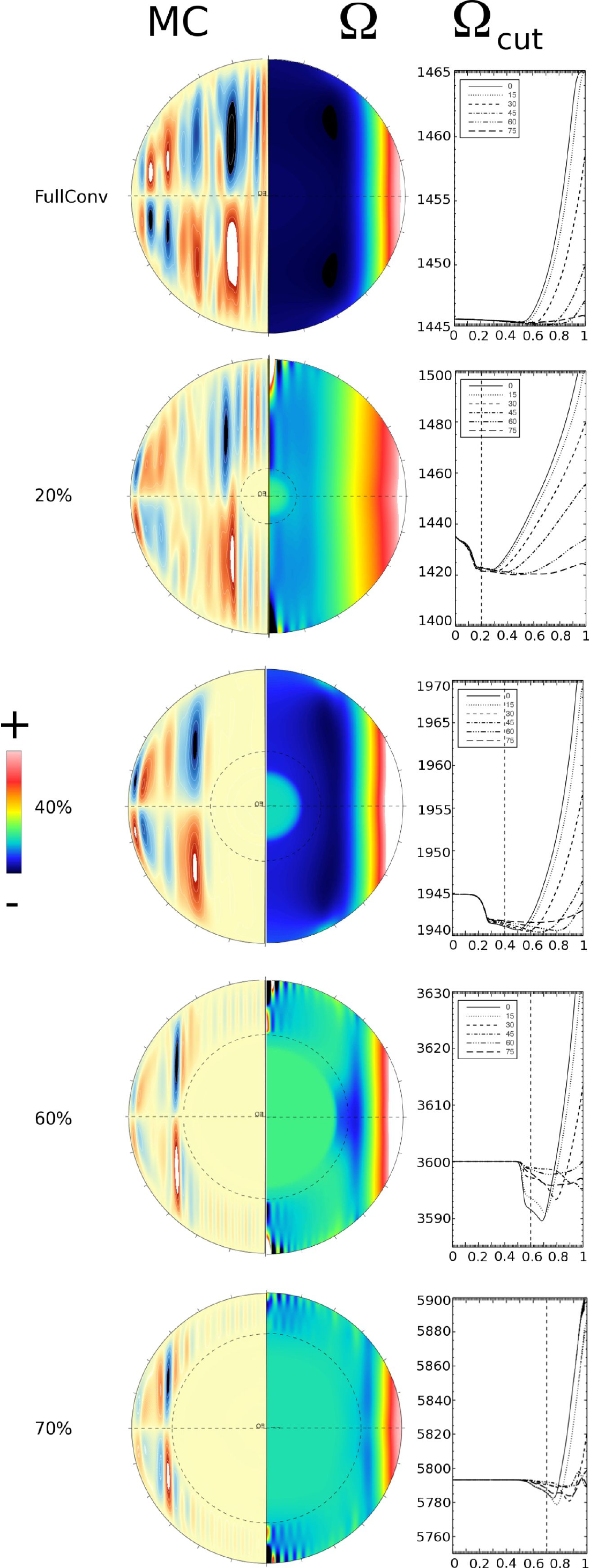}
    \caption{\label{DiffRot} \textit{Contour plot-left side:} Meridional circulation of HD simulations. 
      Clockwise circulation is represented in red and counter-clockwise in blue.
      \textit{Contour plot-right side:} Differential rotation of HD models: FullConv, 20\%, 40 \%, 60\% and 70\%. 
      These models have different rotation rates: 3.5, 3.5, 4.7, 8.7 and 14 (in solar units). 
      Differential rotation in the five simulations is cylindrical. 
      Moreover like in the Sun, the rotation profile is prograde, i.e. the equator rotates quicker than the poles.
      \textit{1D plots:} Radial cuts of the differential rotation profile.
      The radiative cores, when presents, are in solid-body rotation.
      There is a drop of the rotation rate in the tachocline with respect to the value observed in the radiative core.
    }
  \end{center}
\end{figure}
As the rotation changes, differential rotation profiles also vary. 
Figure \ref{DiffRot} shows the differential rotation profile of our models in the meridional plane, averaged over longitude and 400 days. 
All our models have a solar-like profile with a fast prograde equator and slow retrograde poles which is coherent with their Rossby number \citep[see][]{Gastine2014,Brun2017}. 
They are also solar-like in the sense that their rotation rate monotically decreases from the equator to the poles, except for a small region around the poles where the average is not stable due to small level arm. 
Profiles are more cylindrical than the one provided by helioseismology for the Sun. 
This effect is expected for rapidly rotating stars and linked to Taylor-Proudman constraint \hbox{\citep[][]{Brun2002,Brown2008,Brun2017}}. 
In Figure \ref{DiffRot}, radial cuts show an important radial shear at low latitudes. 
\begin{table}
\begin{center}
\caption{Contrasts in differential rotation}\label{omega}
\vspace{0.2cm}
\begin{tabular}{cccc}
\tableline
\tableline
Case     & $\Delta \Omega_{\rm{lat}}$ & $\Delta \Omega_r$ & $\Delta \Omega_{\rm{lat}} / \Omega_0$ \\ [0.5ex]
\tableline
\tableline
FullConv & 18.0                       & 19.4              & $1.2 \times 10^{-2}$                            \\ 
20 \%    & 75.1                       & 86.3              & $5.2 \times 10^{-2}$                            \\
40 \%    & 27.8                       & 30.6              & $1.9 \times 10^{-2}$                            \\
60 \%    & 56.2                       & 67.5              & $3.9 \times 10^{-2}$                            \\
70 \%    & 89.7                       & 96.7              & $6.0 \times 10^{-2}$                            \\ [0.5ex]
\tableline
\tableline
\end{tabular}
\end{center}
\textbf{Note:} Differential rotation in nHz with $\Delta \Omega_{\rm{lat}}$ measured near the surface and $\Delta \Omega_r$ measured at the equator between the surface and the base of the convective zone. 
$\Delta \Omega_{\rm{lat}} / \Omega_0$ represents the relative latitudinal shear measured near the surface. 
All the values have been averaged over 400 days.
\end{table}

Except for the fully convective one, all our models have two zones with a convective envelope and a radiative core. 
Another characteristic of the solar differential rotation profile is the solid body rotation of the radiative core. 
In the first model, with a small radiative core of 20\% of the stellar radius, the core rotation is not constant. 
However is does depend only of the radius and no longer of the latitude. 
In models with a larger core, we see that the core is in solid rotation, except for the \textit{overshooting} zone where the convective motions penetrate the radiative zone. 
There is an interface of shear between the differentially rotating convection zone and the radiative interior which is in solid body rotation. 
This interface is called a \textit{tachocline} \citep[][]{SpiegelZahn1992} and plays an important role in the dynamo process \citep[][]{Browning2006}.

A more quantitative analysis of the differential rotation can be achieved by calculating the following quantities : $\Delta \Omega_{\rm{lat}}$ and $\Delta \Omega_{\rm{r}}$. 
$\Delta \Omega_{\rm{lat}}$ is the contrast in differential rotation near the surface between two latitudes : $0\degree$ and $60\degree$. 
$\Delta \Omega_{\rm{r}}$ is the difference in differential rotation near the equator at two depths : near the surface and at the base of the convective zone. 
The values of these quantities are listed in Table \ref{omega}. 
As the rotation rate of the star increases, from the 40\% model up to the model 70\%, angular velocity contrast increase whereas the relative shear remains quite constant.

As the Rossby number is small in all our simulations, we expect our meridional circulation to be multi-cells which is the case \citep[][]{FeatherstoneMiesch2015}.
Meridional circulation can be represented by plotting the contours of the stream function $\Psi$, as defined by \cite{Miesch2000}: 
\begin{equation}
r \sin \theta \langle \bar{\rho} v_r\rangle = - \frac{1}{r} \frac{\partial \Psi}{\partial \theta} \quad\hbox{and}\quad r \sin \theta \langle \bar{\rho} v_{\theta} \rangle = \frac{\partial \Psi}{\partial r}
\end{equation}
In Figure \ref{DiffRot}, cells in in the stellar interior are cylindrical and aligned with the rotation axis for all models. 
The amplitude of $\rm{v}_{\theta}$ is between 8 and 25 $m.s^{-1}$, depending on the model with no clear trend as the star evolves along the PMS. 
In each hemisphere, there is an changeover of sign between the cylindrical cells with an anti-symmetry with respect to the equator. 
At the surface, the meridional circulation keeps the same behavior, regardless of the model. 
It is counter-clockwise in the northern hemisphere and clockwise in the southern, i.e. the flows go from the equator to the pole 
at the surface and come back to the equator deeper in the star's interior. 

Changes in rotation rate and in geometry of the convective envelop also impact the angular momentum transport. 
Since we choose stress-free and potential-field boundary conditions at the top of our simulations, 
no net external torque is applied, and thus angular momentum is conserved. 
Following previous studies \citep[e.g.][]{Brun2002}, we study the contribution of the different terms in the balance of angular momentum:  
viscous diffusion, meridional circulation and Reynolds stresses.
In all models, Reynolds stresses redistribute the angular momentum outward while the viscous diffusion is inward, down the radial gradient of $\Omega$. 
The amplitude of the contribution of meridional circulation is smaller and its sign varies with radius following the numbers of dominant cells as seen in Figure \ref{DiffRot}. 
Overall the radial balance is well established in the five 3D hydrodynamical simulations. 
Considering the latitudinal flux balance, Reynolds stresses carry angular momentum towards the equator as they are positive (resp. negative) in the northern (resp. southern) hemisphere. 
Viscous diffusion is poleward since they tend to erase the differential rotation in the star. 
Meridional circulation is a response to the torque applied by the sum of the Reynolds and viscous stresses and its sign varies with the multiple cells seen in \ref{DiffRot}. 
Latitudinal balance is longer to attain than the radial one and is not fully established in our hydrodynamical simulations. 

\begin{table*}[ht!]
\begin{center}
\caption{Kinetic energies in hydrodynamical simulations}\label{energies}
\vspace{0.2cm}
\begin{tabular}{cccccccc}
\tableline
\tableline
\\[-1.5ex]
Case  & EK              & KE                     & DRKE                   & MCKE                          & CKE                    & $EK_{RZ/CZ}$              & $KE_{RZ/CZ}$ \\ [0.5ex]
      & ($10^{39}$ erg) & ($10^6$ erg.cm$^{-3}$) & ($10^6$ erg.cm$^{-3}$) & (erg.cm$^{-3}$)               & ($10^5$ erg.cm$^{-3}$) &                           &              \\ [0.8ex]
\tableline
\tableline
      &                 &                        &                        &                    &                        &                           &              \\ [-1.5ex]      
FC    & 114             & 5.52                   & 5.12  (92.7\%)         & 379  $(0.0068 \%)$ & 4.05  (7.30\%)         &  -                        &  -           \\
20 \% & 62.9            & 6.89                   & 6.39  (92.7\%)         & 457  $(0.0066 \%)$ & 5.04  (7.30\%)         & $4.0 \times 10^{-3}$      & 0.5          \\
40 \% & 4.35            & 1.11                   & 0.776 (70.2\%)         & 68.2 $(0.0062 \%)$ & 3.29  (29.8\%)         & $1.5 \times 10^{-2}$      & 0.22         \\
60 \% & 4.04            & 3.11                   & 2.63  (84.5\%)         & 170  $(0.0055 \%)$ & 4.82  (15.5\%)         & $1.2 \times 10^{-2}$      & 0.15         \\
70 \% & 2.01            & 2.18                   & 1.78  (81.1\%)         & 454  $(0.021 \%)$ & 4.11  (18.8\%)         & $2.7 \times 10^{-2}$      & 0.05          \\ [0.5ex]
\tableline
\tableline
\end{tabular}
\end{center}
\textbf{Note:} The first column gives the global energy in the convective zone (in erg).
The four following columns show the kinetic energy density (KE) split into three components : convection (CKE), differential rotation (DRKE) and meridional circulation (MCKE). 
The kinetic energy densities KE, DRKE, MCKE and CKE are reported in erg cm$^{-3}$. 
They take into account the changes in size and geometry of the convective zone in the different models. 
All values are averaged over a period of 400 days.
\end{table*}

\subsection{Kinetic energy}

To further assess the dynamics in the five cases, we now turn to study their energetic content. 
The total energy contained in the convective zone decreases during the PMS as seen in Table \ref{energies}. 
However this trend can be due to the decrease of the size of the convective envelop. 
Thus we look at the kinetic energy density that does not depend on the volume and we notice that there is no clear trend. 
Kinetic energy density can be split into different components linked to the fluctuating convection (CKE), the differential rotation (DRKE) and to the meridional circulation (MCKE) as done by \cite{Miesch2000}. 
These energies are defined as: 
\begin{equation}
\begin{matrix}
\text{DRKE} & = & \displaystyle{\frac{1}{2} \bar{\rho} \langle v_{\phi}\rangle^2 }\\
\text{MCKE} & = & \displaystyle{\frac{1}{2} \bar{\rho}\left(\langle v_r\rangle ^2 + \langle v_{\theta} \rangle^2\right)}\\
\text{CKE}  & = & \displaystyle{\frac{1}{2}\bar{\rho} \left[ \left(v_r- \langle v_r\rangle\right)^2 + \left(v_{\theta} - \langle v_{\theta} \rangle \right)^2 + \left(v_{\phi} - \langle v_{\phi} \rangle\right)^2 \right]} \\
\end{matrix}
\end{equation}
with KE = DRKE + MCKE + CKE and $\langle \cdot \rangle$ is the longitudinal average. 
The values corresponding to these energy densities are reported in Table \ref{energies}. 
In the convective zone, the MCKE is negligible compared to CKE and DRKE. 
The differential rotation has the more important contribution in all models even if this proportion can vary, from 70\% in the 40\% model to 92.7\% in the 20\% model. 
The fraction of the fluctuating convection is smaller but not negligible. 

In the radiative zone, proportions are drastically different since the core is stably stratified. 
Convection and meridional circulation are quite reduced and hence the associated energy densities are negligible compared to the one due to differential rotation. 
DRKE represents more than 99\% of the kinetic energy of all the models possessing a radiative core. 
We defined two quantities to study the evolution of kinetic energy in both zones: $EK_{RZ/CZ}$ is the ratio between the kinetic energy  
and $KE_{RZ/CZ}$ is the ratio between the kinetic energy densities. 
The values of these ratios are given in Table \ref{energies}. 
Hence we notice two opposite trends. 
In total value, the energy in the radiative zone tends to increase more rapidly than the one in the convective zone, 
even if the energy stored in the convective zone is still much higher than the one in the core. 
However the sizes of both zones change since the radiative core grows. 
Thus we also have to look at the ratio of energy densities. 
At first we notice that the fraction of density energy stored in the radiative core is not negligible, especially when the core is small. 
Secondly, this fraction tends to decrease as the radiative core grows. 
To put it in a nutshell, there is more and more energy in the radiative core and its contribution to the total energy, even if it remains small, grows. 
But by looking at the energy densities, we notice that if the energy ratio grows, the energy density ratio decreases as the radiative core grows:  
there is more energy in the radiative zone but it is less concentrated.

\section{Magnetic field properties and evolution during the PMS phase}\label{DynamoAction} 

In the previous section, we saw how the evolution of the star along the PMS modified its internal structure and flows.
We now want to study the impact of this evolution on the resulting internal magnetic field of the star both in the convective and radiative zones. 

\begin{figure}[t]
\begin{center}
    \includegraphics[scale=0.35]{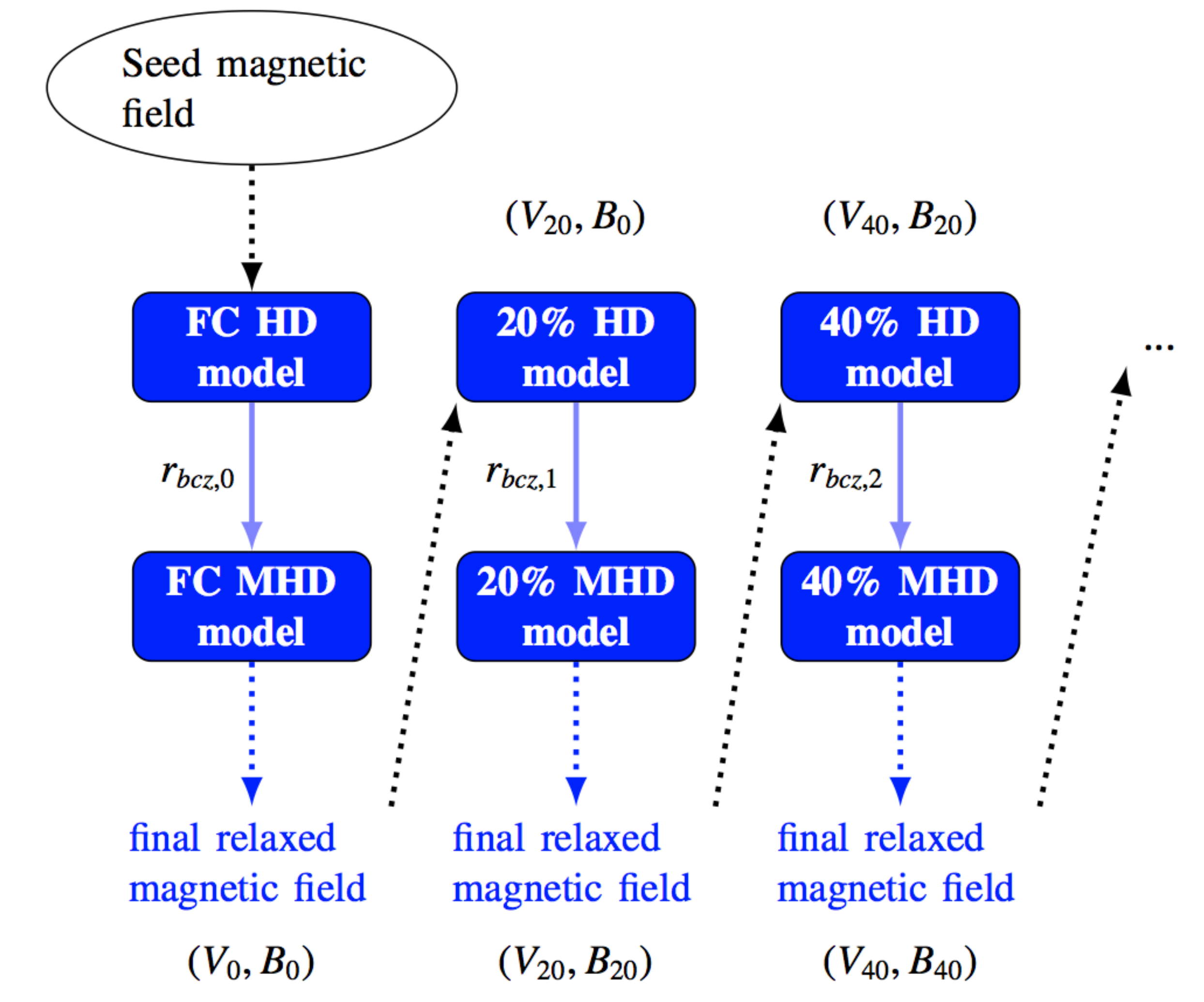}
    \caption{\label{fig::scheme} Description of the procedure to study the evolution of an stellar magnetic field through the PMS.
Once we verified that the hydrodynamical models have equilibrated internal flows and coupling between the radiative core and the convective envelop, 
we introduce a \textit{seed magnetic field} in the first model, here the fully convective one. 
This field is chosen to represent the internal magnetic field of the star after the proto-stellar phase. 
We run the computation of the fully convective model with this magnetic field until this model is equilibrated, we take its final magnetic field and put it in the following model. 
Then we re-do all the steps until we reach the end of the PMS with a star with a 70\% radiative core.}
\end{center}
\end{figure}

\subsection{The procedure}

\begin{figure}[t!]
  \begin{center}
    \includegraphics[scale=0.4]{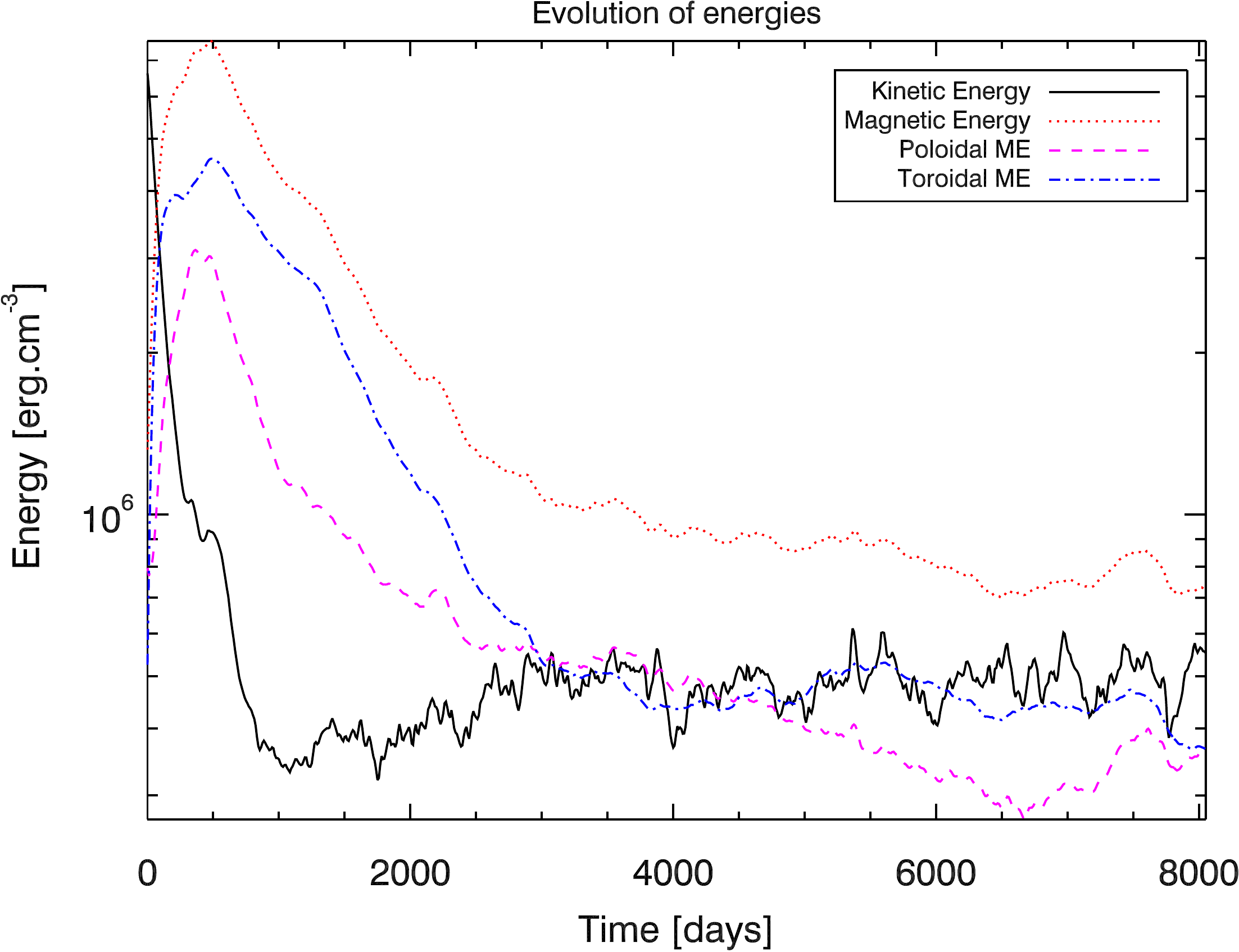}
    \caption{\label{Scalar_mag} Evolution of energies when magnetic field resulting from the fully convective MHD model is introduced into the hydrodynamical simulation of the 20\% model. 
 }
  \end{center}
\end{figure}

Figure \ref{fig::scheme} shows how we proceed to reproduce this evolution with ASH simulations. 
First of all, we inject a seed magnetic field in the fully convective hydrodynamical model. 
This weak seed confined dipole magnetic field represents the field left by the proto-stellar phase. 
We run the MHD simulation of the fully convective model until it reaches an equilibrium state with a dynamo generated field. 
Then we inject the magnetic field resulting from this simulation into the 20\% hydrodynamical model. 
Hence, we can see how the change of internal structure affects the magnetic field. 
Once this simulation reaches an equilibrium state, in the statistically stationary sense (see Figure \ref{Scalar_mag}), and the magnetic field has relaxed in the radiative core, we introduce the resulting magnetic field in the following hydrodynamical model. 
By reproducing these operations with all the hydrodynamical models, we can see the influence on the magnetic field of the changes of internal structure and rotation rate caused by stellar evolution.

\subsection{Magnetic field generation and evolution}

\begin{figure*}
  \begin{center}
    \includegraphics[scale=0.66]{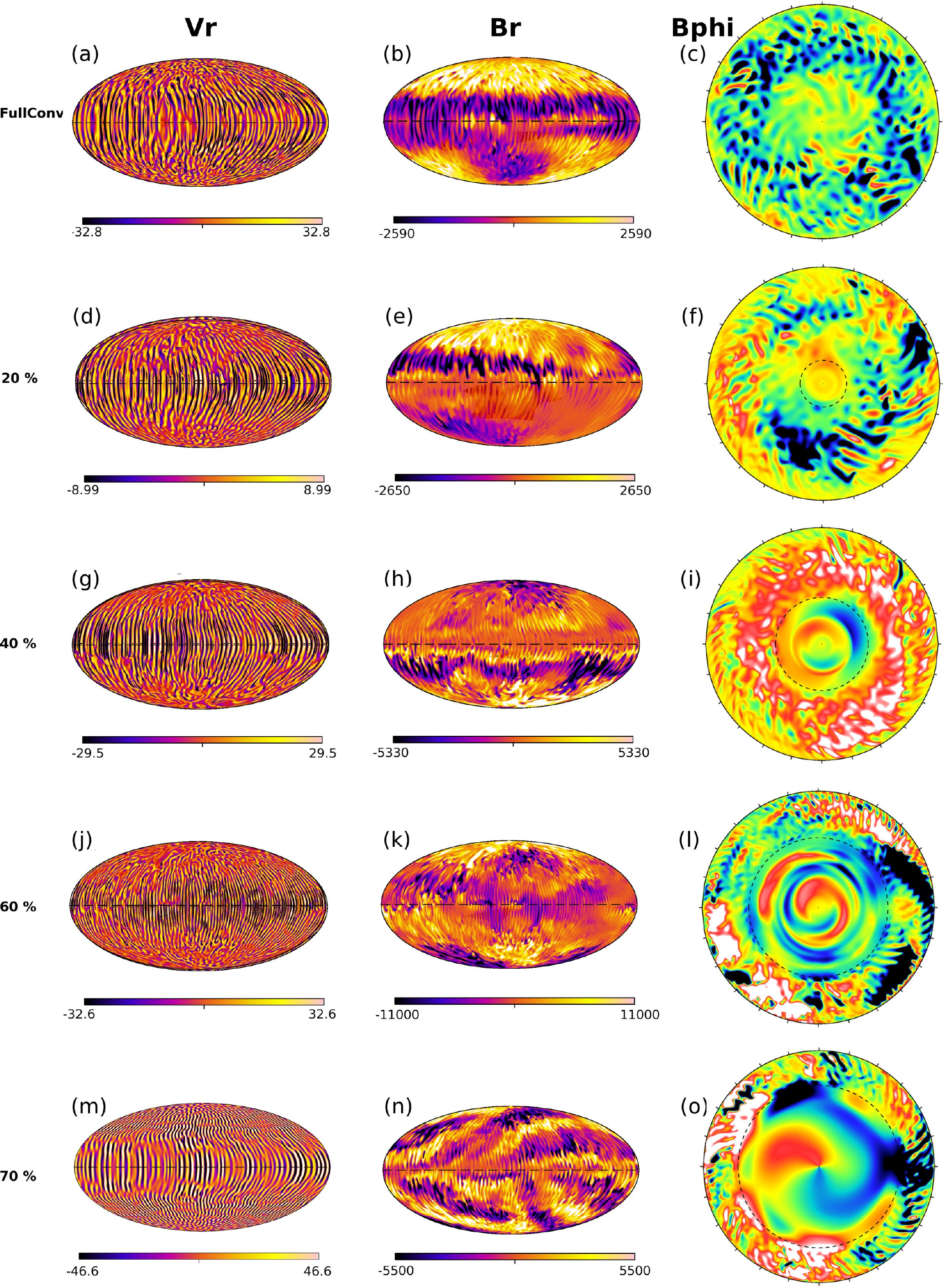}
    \caption{\label{SHSL_Comp}The two first columns show the shell slices at the surface of radial velocity and radial magnetic field for the five MHD simulations. 
      The last column shows the equatorial slice of $B_{\varphi}$ for each model. }
  \end{center}
\end{figure*}

\begin{figure}
  \begin{center}
    \includegraphics[scale=0.5]{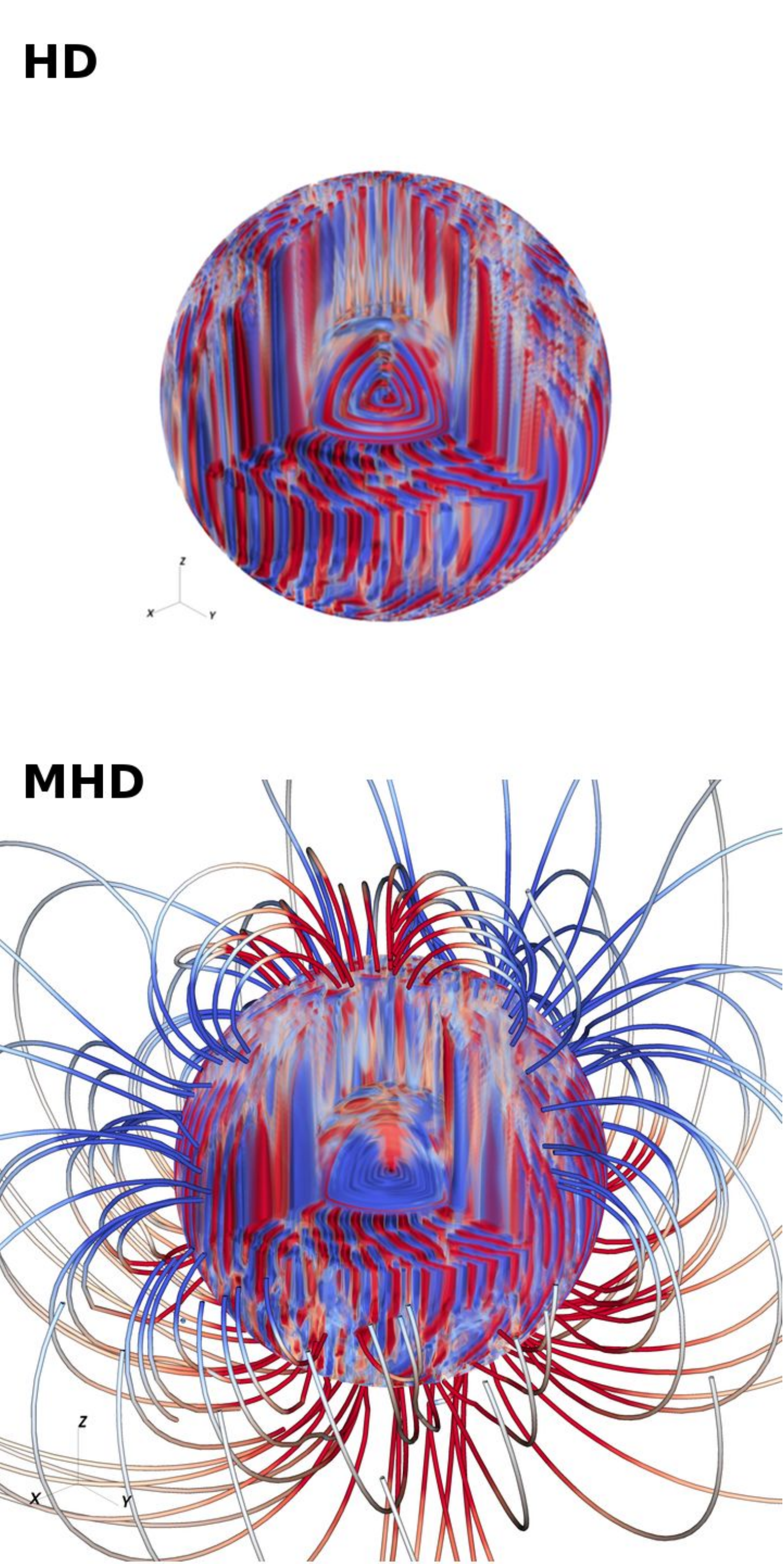}
    \caption{\label{3D}3D views of the 40\% model.
      \textit{Top: } Radial velocity of the hydrodynamical model. 
      \textit{Bottom: } Radial velocity of the MHD model with the potential extrapolation of magnetic field outside the star.
      Upflows in red and downflows in blue. 
      Field lines are color coded with the radial component of the magnetic field. 
   }
  \end{center}
\end{figure}

\begin{figure}
  \begin{center}
    \includegraphics[scale=0.4]{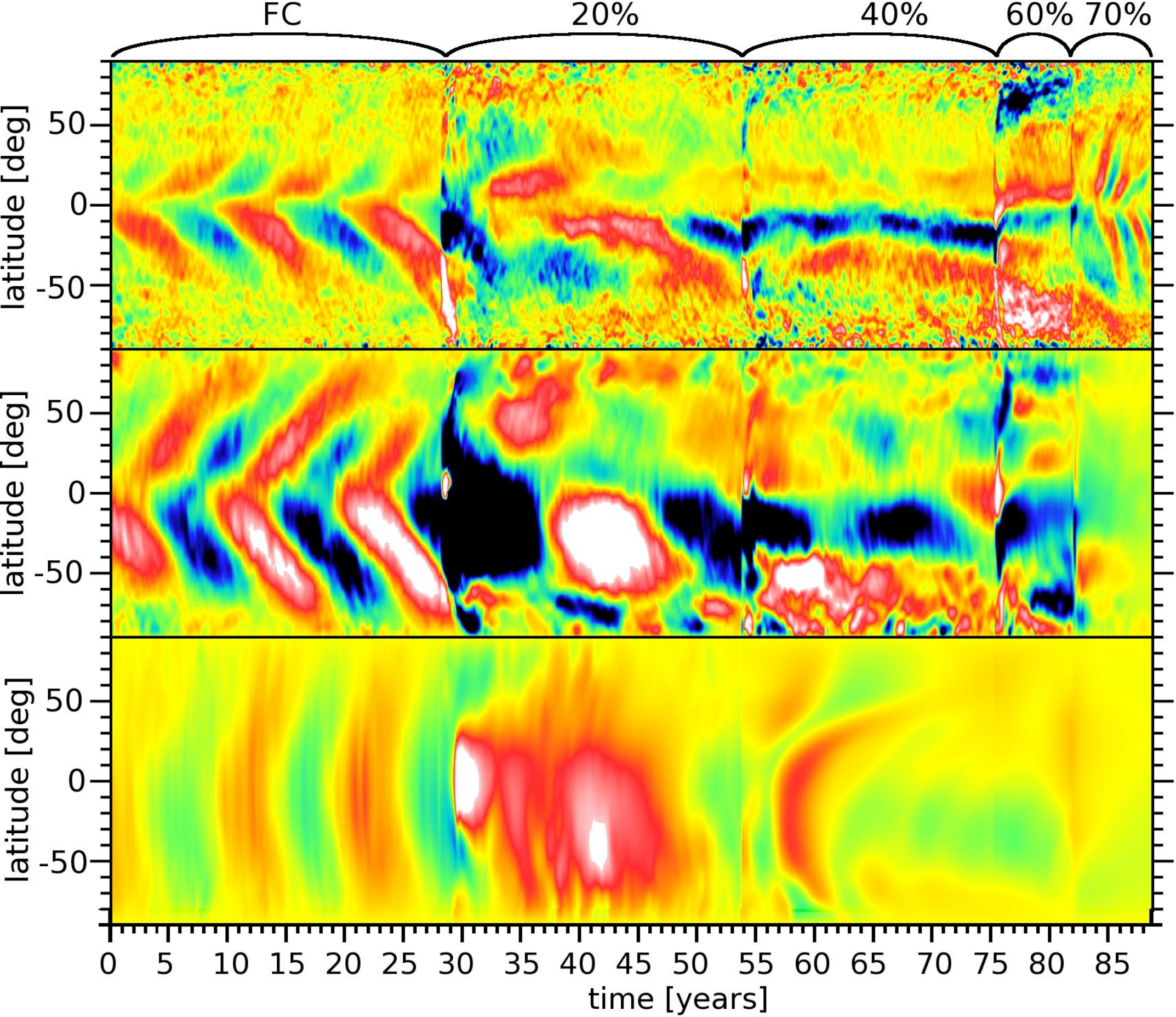}
    \caption{\label{fig:Butterfly}Butterfly diagram for the longitudinal component of the magnetic field $\langle B_{\varphi}\rangle$ across the PMS. 
The five cases are represented on this 2D plot in time and latitude at different radii of 96\%, 60\% and 20\% of the stellar radius (for each model). 
As we go from one model to another, changing the internal structure but propagating the magnetic field, we can see that structures are conserved.}
  \end{center}
\end{figure}

Magnetic and velocity fields are linked by dynamo action and Lorentz forces. 
Hence, as we injected the resulting magnetic field of a model $n$ into the following hydrodynamical model $n+1$, this field has to adapt to the new internal structure and flows. 
Moreover, the injection of the magnetic field also has an influence on the internal structure and flows. 
Figure \ref{Scalar_mag} shows the evolution of the kinetic and magnetic energies in the 20\% model after the injection of the magnetic field resulting from the fully convective dynamo model as we now run it in MHD mode.
Kinetic energy drops with the presence of the magnetic field. 
The decrease is due to a drastic change in the differential rotation profile which will be describe below (see section \ref{Comp_HD_MHD}).
Magnetic energy has a burst as it has to adapt to the new internal structure and flows. 
Both poloidal and toroidal energies grow before decreasing. 
After a transient phase, of roughly 2500 days, kinetic and magnetic energies stabilize and a genuine dynamo process occurs in the convective envelop. 

Figure \ref{SHSL_Comp} shows us radial velocity at the surface of the stars. 
By comparing it with those observed in \ref{sec::HD}, we see similar patterns with prograde banana cells at the equator. 
Size of convective patterns do not seem to change much between the HD and MHD simulations. 
The 3D topology of the radial velocity is illustrated by for the 40\% model Figure \ref{3D} both in HD and MHD cases. 
We notice the cylindrical patterns of convection linked to fast rotation in both cases. 
The tangent cylinder can be seen as velocity is smaller in it. 
These two figures, \ref{SHSL_Comp} and \ref{3D}, also show the topology of the magnetic field inside and outside the star. 
The radial component of the magnetic field is shown at the surface of the star and we see well-defined patterns with a growing amplitude as the star grows along the PMS, 
except for the 70\% which has an amplitude of $B_r$ similar to the 40\% model. 
$B_{\varphi}$ is shown with equatorial slices. 
In these slices we can notice that there is two different behaviors of magnetic field depending on the zone, convective or radiative, of the star. 
In the convective zone, the magnetic field, that comes from the dynamo process, is quite turbulent whereas in the radiative core the field relaxes and possesses smoother and larger structures. 
The 3D view shows us a potential extrapolation of the magnetic field outside the star which is complex, highly non-axisymmetric and exhibits as well extended transequatorial loops.

As we propagate the magnetic field from one model to another, we want to analyze its time evolution through the PMS. 
Thus we plot, in Figure \ref{fig:Butterfly}, a butterfly diagram of our complete set of simulations. 
This diagram is a 2D plot in time and latitude at different radii (96\%, 60\% and 20\% of the stellar radius) of $\langle B_{\varphi} \rangle$, the longitudinal magnetic field averaged over $\varphi$. 
At 96\% of the stellar radius, i.e. at the surface of the simulation, and at 60\% of the stellar radius, we see cycles in the fully convective and 20\% models (see the upper and middle panels). 
A finer study of these cycles will be done in section \ref{cycles} on the cycle period and the sense of the dynamo wave. 
Sharp transitions occurs as the magnetic field goes from one model to the following, which is coherent with the burst observed in the volume integrated energy analysis. 
However some magnetic structures are preserved even during the transition between the different simulations. 
In the lower plot, at 20\% of the stellar radius, we observe that the butterfly diagram has a different behavior depending on the models. 
In the fully convective star, we still see cycling patterns and a propagation of the field from the equator to the poles which is coherent since we are still in the convective zone of the star. 
When the magnetic field is injected into the 20\% model, there is a major change. 
Indeed, in this model, the observed radius is no longer in the convective envelop, but in the tachocline of the star. 
We can notice that we do not see any propagation patterns and the amplitude of $B_{\varphi}$ is higher than in the convective case, 
likely due to the larger radial shear of $\Omega$ in the 20\% model. 
This can be explain as the tachocline is a zone of high shear where global dynamo might be seated. 
From the 40\% model to the 70\% one, the observed radius lies in the radiative core. 
The amplitude is much lower than in the tachocline and less structured than in the fully convective model, i.e. we see not cycling patterns. 
The magnetic field relax in the radiative zone until the ZAMS. 

Hence we see that $\BB$ evolves quite substantially from one model to the other. 
The source of these changes, and the type of field generation it produces, will be discussed in section \ref{ExplainingTheDynamics}. 

\begin{figure}[h!]
  \begin{center}
    \includegraphics[scale=0.71]{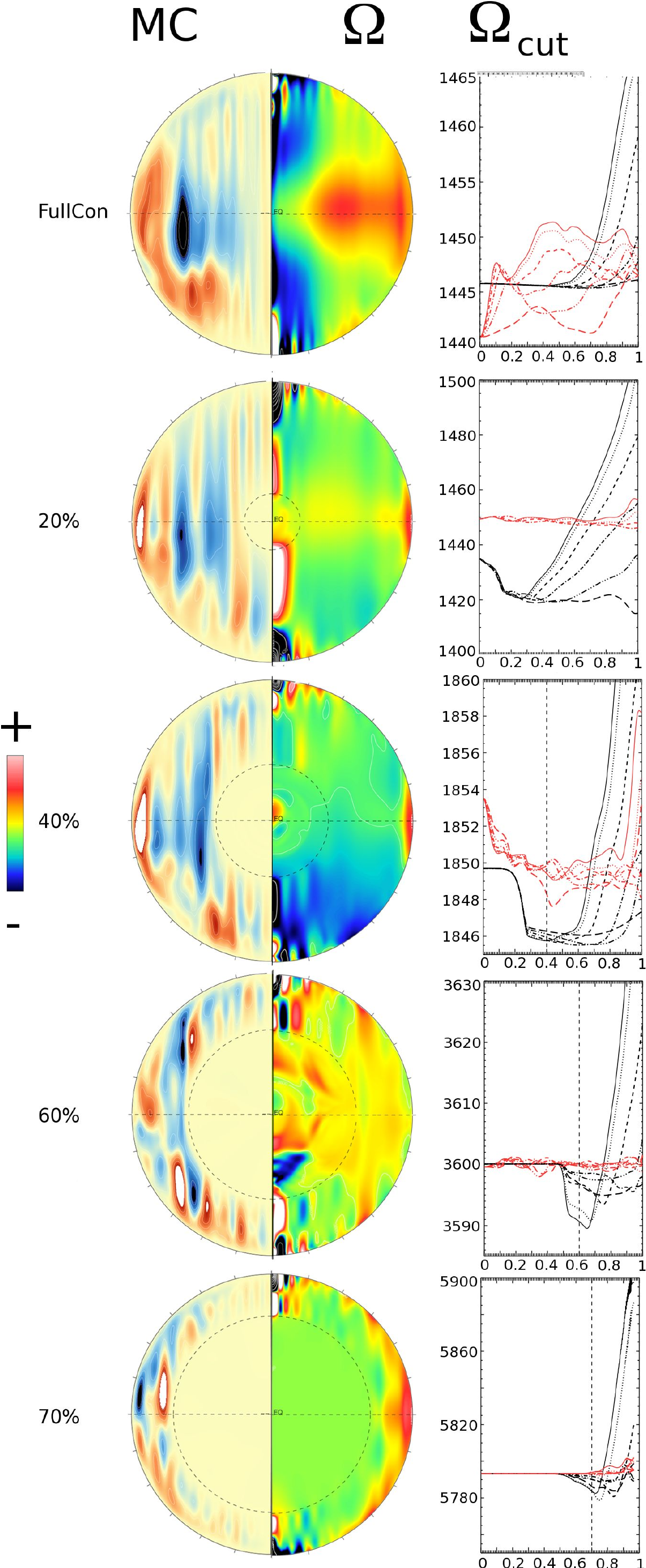}
    \caption{\label{DiffRot_Comp}\textit{Contour plot-left side:} Meridional circulation of the MHD simulations.
    \textit{Contour plot-right:} Differential rotation of the MHD simulations.
    \textit{1D plots}: Radial cuts of the differential rotation profile. Black plots come from the HD simulations and the red plots results from the MHD models. 
    In all models, the differential rotation is flatten by the introduction of magnetic field.}
  \end{center}
\end{figure}

\subsection{Mean flows HD vs. MHD \label{Comp_HD_MHD}}

\begin{figure}[t!]
  \begin{center}
    \includegraphics[scale=0.45]{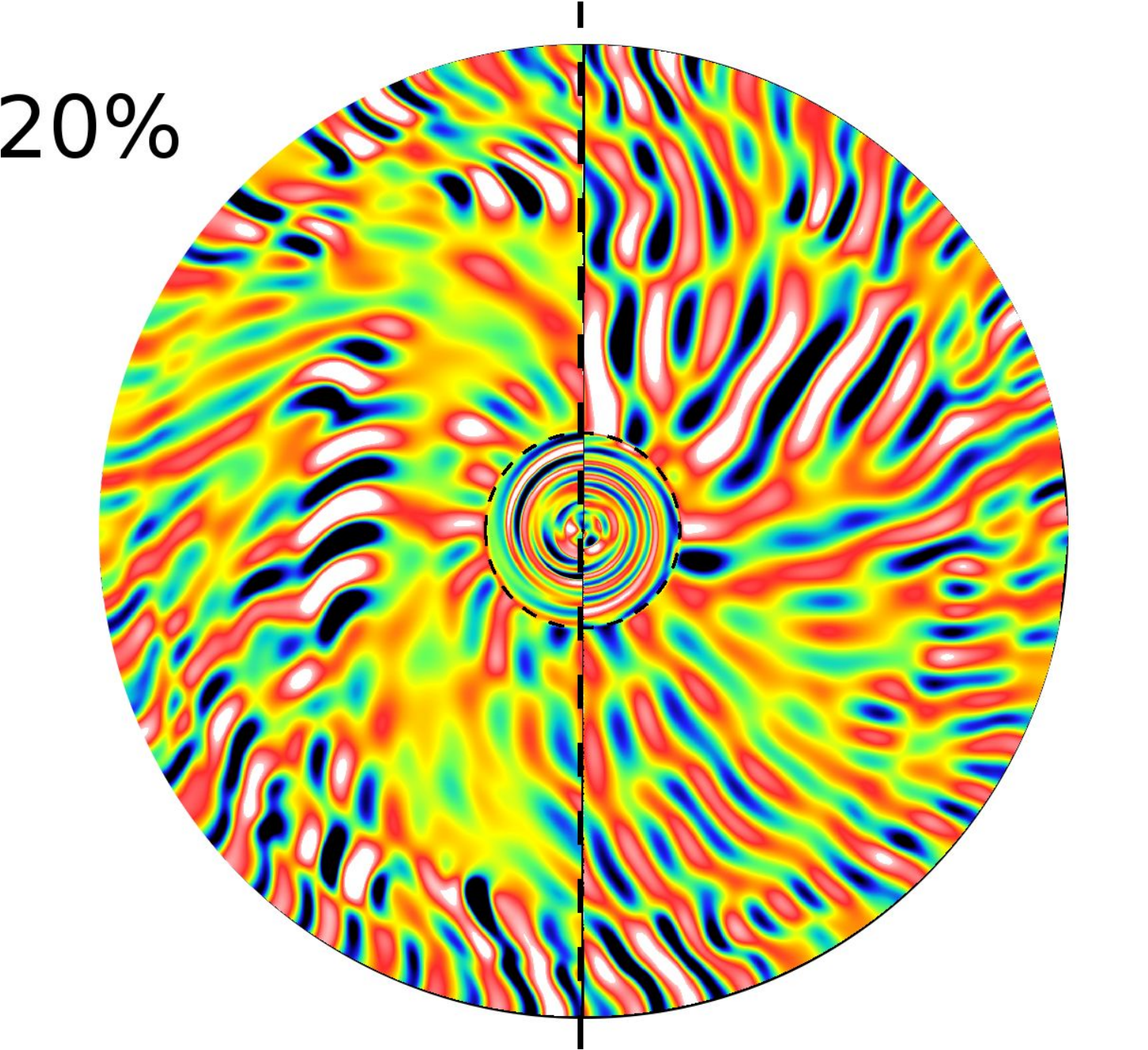}
    \caption{\label{EQSL_Comp}As in Figure \ref{EQSL}, we show here the equatorial slices of normalized radial velocity for the 20\% HD and MHD simulations, 
    with the downflows represented in blue/dark and upflows in red/white. 
    The HD simulation is shown on the left side while the MHD one is plotted on the right side of the slice. 
    We see that the internal magnetic field does not drastically change the horizontal extent of the convective patterns whereas the radial patterns are larger in the MHD simulation than in the HD one.}
  \end{center}
\end{figure}

The introduction of magnetic field in the hydrodynamical models strongly impacts the internal flows we studied in the previous section \ref{sec::HD}. 
One major change is the profile of differential rotation due to the influence of Maxwell stresses \citep[][]{Brun2004}. 
Comparison between hydrodynamical and MHD simulations are shown in Figure \ref{DiffRot_Comp}. 
We can see that the presence of magnetic field tends to reduce the latitudinal variation of the differential rotation in the convective envelop. 
Solid rotation in the radiative core is also altered by magnetic fields. 
This change is certainly due to diffusion and spreading of the differential rotation of the convective envelope into the radiative zone. 
The rotation profiles of the MHD simulations remain prograde but less monotonic and contrasted. 
Changes on structures of the internal flows can be observed through the equatorial slice of the 20\% model in Figure \ref{EQSL_Comp}. 
In this figure, the left side of the slice comes from the progenitor HD simulation and the right side shows the result of the MHD model. 
The horizontal extent of the convective patterns do not drastically change when the magnetic field is introduced. 
On the contrary, we notice that the radial extent of these structures is larger in the MHD models that in the hydrodynamical ones.
This property is coherent with the changes in differential rotation profiles shown in Figure \ref{DiffRot_Comp} and can be seen in each simulation. 
Indeed, as the differential rotation profiles are flatter in the MHD models, the shearing is smaller and thus the radial extent of the convective structures are more elongated. 

Meridional circulation is also impacted by the introduction of the magnetic field. 
The cells are still cylindrical and aligned with the rotation axis but there is no anti-symmetry at the equator, as it was observed in the hydrodynamical case. 
A large clockwise cell spread both sides of the equator for the first two models. 
In the simulations with a larger core, this cell breaks and we see two smaller cells with opposite signs on both sides of the equator. 

The change in differential rotation, observed in the MHD simulations, 
can be understood by the presence of two additional contributions to the way angular momentum is redistributed 
in the convective shell: e.g. Maxwell stresses and large scaled magnetic torques \citep[][]{Brun2004}.
As seen in section \ref{sec::HD}, Reynolds stresses carry angular momentum outward whereas the viscous diffusion is inward. 
The introduction of the magnetic field modifies this balance as the inward transport in no longer supported by the viscous diffusion, 
since the differential rotation has been quenched, but by the Maxwell stresses and large scale magnetic torques. 
In the MHD simulations, the radial balance of angular momentum transport is well established. 
However the latitudinal balance is not fully achieved yet when we propagate the magnetic field from one model to the following. 
We choose to stop our simulations when DRKE is mostly constant in time to inject the resulting magnetic field in the following model. 
The latitudinal balance is quite long to established and for computional resources issues we cannot compute each model until it reaches this balance. 
However we notice some trends in the latitudinal transport of angular momentum. 
The Reynolds stresses are still equatorward and the viscous diffusion poleward. 
The transport linked to Maxwell stresses and large scale magnetic torques is not completely stable but is mainly poleward. 
The meridional circulation that helps to established the balance varies in signs as it has multiple cells in each hemisphere as seen in Figure \ref{DiffRot_Comp}. 
Overall, the action of the magnetic field is to quench the angular velocity by reducing both the radial and latitudinal contrast.

\section{Explaining the dynamics}\label{ExplainingTheDynamics}

\subsection{Energy content and radial flux balance\label{KEMHD}}

Kinetic energies are also impacted by the injection of the magnetic field in the model.
By comparing the energy densities between the hydrodynamical models and the MHD ones, we note that there is small variation, between 3 and 12\% (see Table \ref{Kenergies}). 
Contrary to the HD simulations, the energy densities are quite the same in all MHD models. 
Since differential rotation is flatten, the relative influence of the components of the kinetic energy in the convective envelope is changed, as shown in Table \ref{Kenergies}. 
\begin{table*}[t]
\begin{center}
\caption{Kinetic energies of the MHD simulations}\label{Kenergies}
\vspace{0.2cm}
\normalsize
\begin{tabular}{ccccccc}
\tableline
\tableline
\\[-1.5ex]
Case & EK          & KE       & $\Delta$KE & DRKE         & MCKE           & CKE          \\ [0.5ex]
     & ($10^{38}$) & ($10^5$) & (\%)       & ($10^4$)     & ($10^2$)       & ($10^5$)     \\ [0.8ex]
\tableline
\tableline
     &             &          &            &              &                &              \\[-1.5ex] 
FC   & 101         & 4.93     & 12         & 4.0  (8.2\%) & 2.2 ($0.04\%$) & 4.5 (91.8\%) \\
20\% & 45.2        & 4.95     & 4          & 2.5  (5.1\%) & 5.1 ($0.1 \%$) & 4.7 (94.8\%) \\
40\% & 19.4        & 4.95     & 8          & 2.6  (5.2\%) & 4.6 ($0.09\%$) & 4.7 (94.7\%) \\
60\% & 5.74        & 4.42     & 3          & 0.44 (1.0\%) & 2.3 ($0.05\%$) & 4.4 (98.9\%) \\
70\% & 4.17        & 4.51     & 5          & 1.2  (2.6\%) & 2.8 ($0.06\%$) & 4.4 (97.3\%) \\ [0.5ex]
\tableline
\tableline
\end{tabular}
\end{center}
\textbf{Note:} The first column gives the global kinetic energy (EK) of the convective zone (in erg). 
The third column represents the variations of kinetic energy density (KE) between the hydrodynamical and MHD simulations: 
$\Delta$ KE = $|$KE$_{\rm{MHD}}-$KE$_{\rm{HD}}$$|/$KE$_{\rm{MHD}}$
The kinetic energy density is split into three contributions: convection (CKE), differential rotation (DRKE) and meridional circulation (MCKE) for MHD models. 
All energy densities are averaged over a period of 400 days and reported in erg cm$^{-3}$. 
\end{table*}
\begin{table*}
\begin{center}
\normalsize
\caption{Magnetic energies}\label{Menergies}
\vspace{0.2cm}
\begin{tabular}{ccccccc}
\tableline
\tableline
\\ [-1.5ex]
Case  & EM          & ME       & $\frac{ME}{KE}$ & TME          & PME          & FME          \\ [0.5ex]
      & ($10^{38}$) & ($10^5$) &                 & ($10^4$)     & ($10^5$)     & ($10^5$)     \\ [0.8ex]
\tableline
\tableline
      &             &          &                 &              &              &              \\ [-1.5ex] 
FC    & 116         & 5.62     & 1.14            & 11 (19.9\%)  & 1.5 (27.0\%) & 3.0 (53.1\%) \\
20\%  & 68.4        & 7.49     & 1.51            & 6.5 (8.74\%) & 1.1 (14.4\%) & 5.8 (76.9\%) \\
40\%  & 42.0        & 10.7     & 2.16            & 9.5 (8.86\%) & 2.5 (22.9\%) & 7.3 (68.2\%) \\
60\%  & 27.3        & 21.0     & 4.75            & 7.3 (3.48\%) & 3.2 (15.5\%) & 17  (81.0\%) \\
70\%  & 8.13        & 8.8      & 1.95            & 1.2 (1.37\%) & 0.5 (5.23\%) & 8.2 (93.4\%) \\ [0.5ex]
\tableline
\tableline
\end{tabular}
\end{center}
\textbf{Note:} The first column gives the global magnetic energy (EM) of the convective zone (in erg). 
The third column represents the ratio of magnetic to kinetic energy.
The magnetic energy density (ME) is split into three contributions: poloidal mean energy (PME), toroidal mean energy (TME) and fluctuating energy (FME). 
All energy densities are averaged over a period of 400 days and reported in erg cm$^{-3}$.
\end{table*}

In the convective envelop of the MHD simulations, the DRKE decreases compared to the hydrodynamical case and the dominant term becomes the convective one (CKE). 
As in the hydrodynamical models, the contribution of MCKE remains negligible. 

As seen by plotting the temporal evolution of the energy densities, in Figure \ref{Scalar_mag}, the magnetic energy varies when the magnetic field is propagated from one simulation to the following one. 
Thus our MHD models enable us to study how magnetic energy evolves as the magnetic field is propagated along the PMS. 
First of all, Table \ref{Menergies} shows that magnetic energy grows slightly as the star evolves through the PMS. 
For a finer analysis, as for the kinetic energy, we split the magnetic energy (ME) into three different components linked to the mean toroidal magnetic energy (TME), to the mean poloidal magnetic energy (PME) and to the fluctuating energy (FME): 
\begin{equation}
\begin{matrix}
\text{TME} & = & \displaystyle{\frac{1}{8\pi} \langle B_{\varphi} \rangle^2} \\
\text{PME} & = & \displaystyle{\frac{1}{8\pi} \left( \langle B_r \rangle^2 + \langle B_{\theta} \rangle^2 \right) }\\
\text{FME} & = & \displaystyle{\frac{1}{8\pi} \left[ \left( B_r - \langle B_r \rangle \right)^2 + \left( B_{\theta} - \langle B_{\theta} \rangle \right)^2 + \left(B_{\varphi} - \langle B_{\varphi} \rangle \right)^2 \right]}\\
\end{matrix}
\end{equation}
where $\langle \cdot \rangle$ is the longitudinal mean and ME = TME + PME + FME. 

By looking at Table \ref{Menergies}, we note that all models are in a superequipartition state with $ME/KE > 1$. 
The ratio even almost reaches 5 for the 60\%.
Hence in all our models is in an equirepartition state, and even in a superequipartition state for the faster cases.
$ME/KE$ increases as the star ages. 
Only the 70\% model behaves differently with $ME/KE$ decreasing. 
As seen previously, the kinetic energy density does not change much in the MHD models. 
The change observed in the ratio $ME/KE$ is thus mostly due to the change in magnetic energy density. 
Indeed, we observe that $ME$ increases as the star goes along the PMS except for the 70\% model. 
The specificity of this simulation is a change in the stellar luminosity evolution. 
The size of the radiative core and the rotation rate monotically change during the PMS 
whereas the luminosity first decreases down to 60\% model and then increases until the ZAMS. 
This can explain a different behavior for the 70\% model compared to the other four cases. 
Indeed, $ME$ increases with the size of the radiative core and the rotation rate. 
By looking at Table \ref{Menergies}, we note that, in all simulations, magnetic energy in the convective zone is mainly contained in the fluctuating part (FME). 
The proportion of mean field energy (TME and PME) varies strongly as the star evolves along the PMS. 
At the very beginning of the PMS, it represents 47\% of the total energy quite close to the proportion of 49\% found by a study led by \cite{Brown2011}. 
As the star ages, this proportion decreases until 19\% for the 60\% model (similar to results obtained by \cite{Browning2008}). 
Finally as the stars arrives on the ZAMS, mean field energy only represents a few percent of the total energy as in the study led by \cite{Brun2004}. 
However, in all these simulations, the mean toroidal energy prevails in the mean energy, whereas in our models, the predominant term is the mean poloidal energy. 
In summary, the decrease of the mean field energy is coherent with results obtained by \cite{Gregory2012} in which 
the magnetic field becomes less axisymmetric and more complex as the star ages along the PMS.

\subsection{Topology of the magnetic fields \label{sec::topo}}

In the previous section, we have studied the evolution of the amplitude of the magnetic field during the PMS 
and found that ME mostly grows. 
We will now focus on the topology of these fields. 
We can express the energy of the magnetic field at the surface $\text{ME}_{surf}$ as: 
\begin{equation}
\text{ME}_{surf} = \sum_{\ell,m} \text{ME}_{\ell}^m Y_{\ell,m}  
\end{equation}
and we can define: 
\begin{equation}
\left\{
\begin{matrix}
\text{ME}_{\ell} & = & \displaystyle{\sum_m \text{ME}_{\ell}^m, \forall \ell} \\
\text{ME}^m & = & \displaystyle{\sum_l \text{ME}_{\ell}^m, \forall m}. \\
\end{matrix}
\right.
\end{equation}
We define two ratios that characterize the topology \citep[][]{Christensen2006}: the dipole field strength:
\begin{equation}
f_{\rm{dip}} = \displaystyle{\frac{\text{ME}_1}{\sum_{\ell=1}^{12} \text{ME}_{\ell}}}
\end{equation}
and the ratio between the axisymmetric and non axisymmetric field:
\begin{equation}
R_{axi} = \displaystyle{\frac{\text{ME}^0}{\sum_{m>0} \text{ME}^m}}. 
\end{equation}
As in \cite{Schrinner2012}, we also defined the local Rossby number of our simulations 
\begin{equation}
R_{o,l} = R_o \frac{\bar{\ell}}{\pi}
\end{equation}
where
\begin{equation}
\bar{\ell} = \sum_{\ell} \ell \frac{\langle\left(\mathbf{v}\right)_{\ell}\cdot\left(\mathbf{v}\right)_{\ell}\rangle}{\langle\mathbf{v}\cdot\mathbf{v}\rangle}
\end{equation}
is the mean harmonic degree, with $\langle \cdot \rangle$ the average over time and radius. 

\begin{figure}
  \begin{center}
    \includegraphics[scale=0.7]{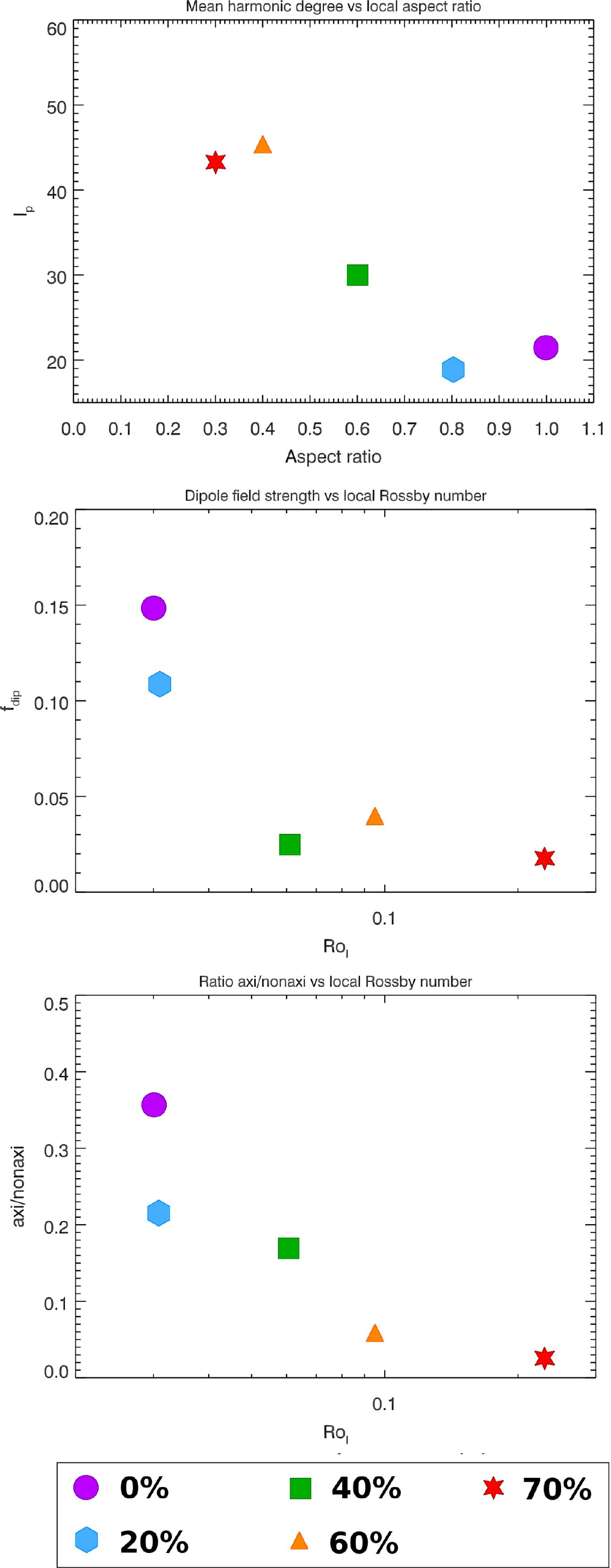}
    \caption{\label{Schr} Evolution of the magnetic field topology with respect to the local Rossby number of our simulations. 
      \textit{Top:} Mean harmonic degree evolution, as the local Rossby number grows, $\bar{\ell}_p$ grows. 
      \textit{Middle:} Dipole field strength decreases as the local Rossby number increases. 
      \textit{Bottom:} The amplitude of the axisymmetric field decreases with respect to the non axisymmetric one as the star evolves along the PMS
}
  \end{center}
\end{figure}

Increasing radiative core and rotation rate lead to a larger mean harmonic degree (see Figure \ref{Schr}). 
This result can be understood by considering the geometrical change of the convective zone, e.g. the smaller aspect ratio that favors larger $\ell$.
As seen in section \ref{setup}, the Rossby number also grows as the star evolves along the PMS. 
This result may seem counterintuitive since as faster rotation should lead to smaller Rossby number. 
However in our cases, rotation rate is not the only stellar parameter to change: the thickness of the convective envelop decreases along the PMS.
By looking at the product $\Omega_0 D$, we see that it decreases as the star ages. 
Hence it is logical that the Rossby number increases as the star evolves along the PMS.
As the mean harmonic degree and the Rossby number grow, the local Rossby number increases as the star is aging. 
By plotting $f_{\rm{dip}}$ as a function of the local Rossby number, \cite{Schrinner2012} noted a transition between the dipolar and multipolar mode at $R_{o,l}^c = 0.1$. 
The local Rossby number of our simulations are around this transition value. 
We also observe a transition in the topology evolution of the magnetic field along the PMS. 
As in \cite{Schrinner2012}, dipolar components of the magnetic field are weak when $R_{o,l} > 0.1$. 
These components are bigger when $R_{o,l} < 0.1$ even if the transition is less strong than the one observed by Schrinner. 
The amplitude of the axisymmetric part of the magnetic field also decreases as the star evolves along the PMS. 
We see a transition between the fully convective model, with $R_{axi} \simeq 0.35$, the two models with a small radiative core, with $R_{axi} \simeq 0.2$, and the two simulations with a big radiative core $R_{axi} < 0.05 $. 
These observations are coherent with the different behavior studied by \cite{Gregory2012}. 
By looking to the dipolar field strength, we also find similarities with this study since dipolar components of the fully convective models are greater than in the other simulations. 
However it has to be tempered since they are still quite weak contrary to what was observed in the study of \cite{Gregory2012}.
Moreover the dipolar field strength is quite different between the two models with a small core even if they are both supposed to have weak dipolar components. 

\begin{figure*}[t!]
  \begin{center}
    \includegraphics[scale=0.5]{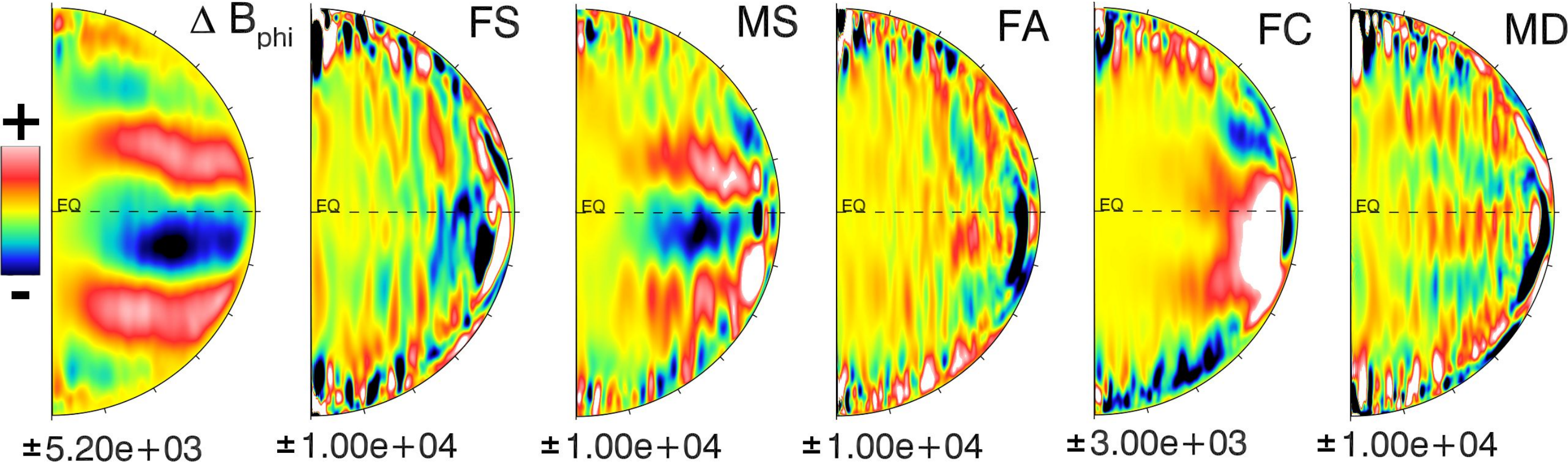}
    \caption{\label{MagFieldProd}Balance between the variation of $B_{\varphi}$ and generation terms of the toroidal magnetic field for the MHD simulation of the fully convective model. 
The first 2D plot shows the variation of $B_{\varphi}$ : $\Delta B_{\varphi}  = \left(B_{\varphi,2}-B_{\varphi,1}\right)$. 
The other plots are the generation terms integrated between $t_1$ and $t_2$. 
All the terms are reported in G/s.
}
  \end{center}
\end{figure*}
\begin{figure}[t!]
  \begin{center}
    \includegraphics[scale=0.4]{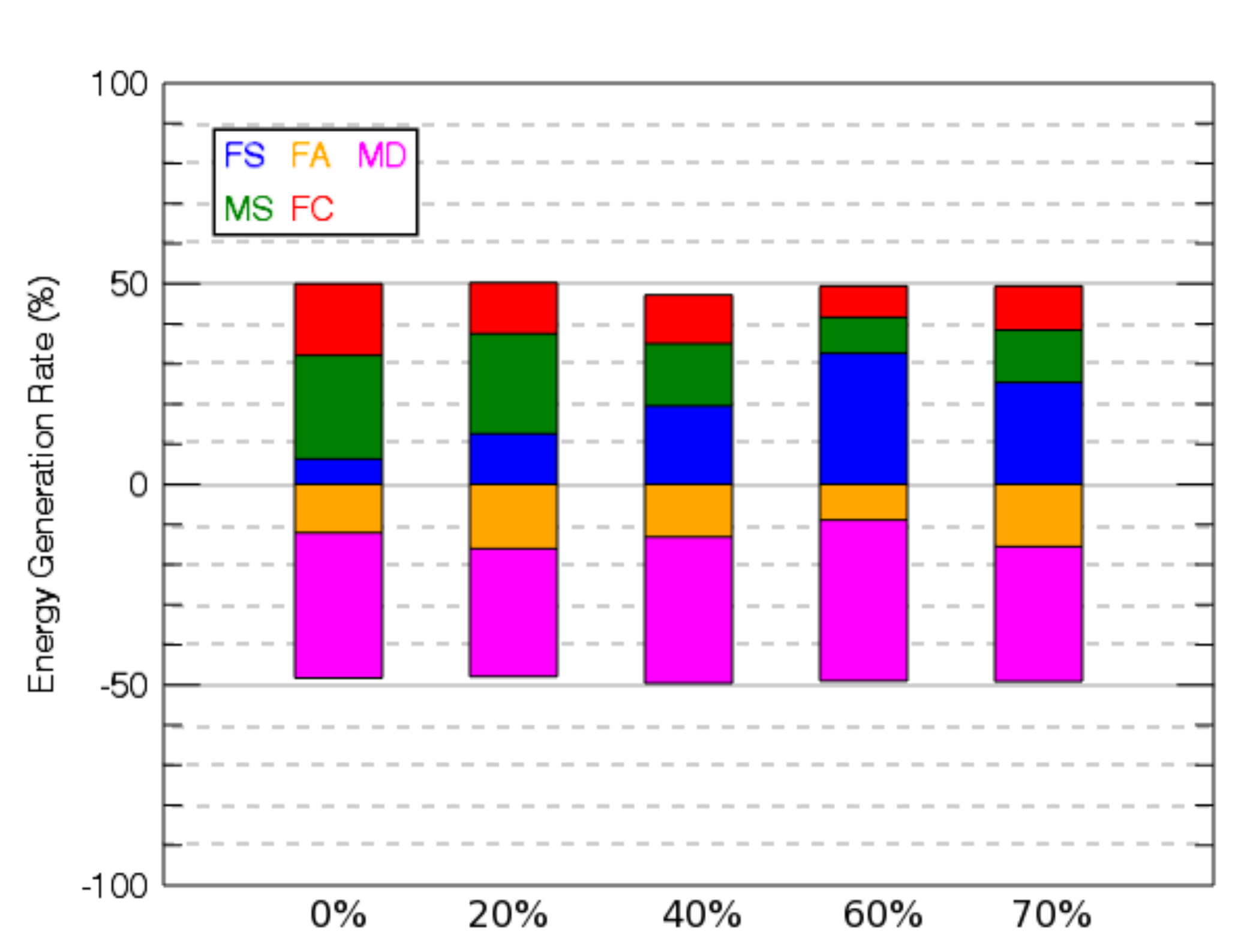}
    \caption{\label{bar}Balance of time-averaged generation terms of the toroidal magnetic energy reported in \%. 
      The toroidal field is maintained thanks to the fluctuating compression and to the shear action, both mean and fluctuating. 
      It is destroyed by the fluctuating advection and mean diffusion terms. 
      The mean advection and mean compression terms are not represented since they are negligible compared to the other terms. 
}
  \end{center}
\end{figure}

\subsection{Mean field generation}\label{MFG}

In order to better understand the dynamics of the magnetic fields, we now turn to the generation of mean field. 
To study the generation of both mean toroidal and poloidal magnetic fields, we use the decomposition of the induction equation \ref{eq::induction} developed by \cite{Brown2011}.
Thus we get the different contributions of shear, advection and compression for the magnetic field production: 
\begin{equation}\label{eq::torgeneration}
\frac{\partial \langle \BB \rangle}{\partial t} = P_{\rm{MS}} + P_{\rm{FS}} + P_{\rm{MA}} + P_{\rm{FA}} + P_{\rm{MC}} +P_{\rm{FC}} + P_{\rm{MD}}
\end{equation}
with $P_{\rm{MS}}$ representing the production of field by mean shear, $P_{\rm{FS}}$ production by fluctuating shear, $P_{\rm{MA}}$ advection by mean flows, $P_{\rm{FA}}$ advection by fluctuating flows, $P_{\rm{MC}}$ amplification due to the compressibility of mean flows, $P_{\rm{FC}}$ amplification due to the compressibility of fluctuating flows and $P_{\rm{MD}}$ the ohmic diffusion of the mean fields: 
\begin{equation}
\begin{matrix}
P_{\rm{MS}} & = & \displaystyle{\left( \langle \BB\rangle \cdot \nabla \right) \langle \mathbf{v} \rangle}, \rule[-7pt]{0pt}{20pt}\\
P_{\rm{FS}} & = & \displaystyle{\langle \left( \BB' \cdot \nabla\right) \mathbf{v}'\rangle}, \rule[-7pt]{0pt}{20pt}\\
P_{\rm{MA}} & = & \displaystyle{- \left( \langle \mathbf{v} \rangle \cdot \nabla\right)\langle \BB\rangle}, \rule[-7pt]{0pt}{20pt}\\
P_{\rm{FA}} & = & \displaystyle{\langle\left( \mathbf{v}' \cdot \nabla\right)\BB'\rangle}, \rule[-7pt]{0pt}{20pt}\\
P_{\rm{MC}} & = & \displaystyle{\left(\langle v_r\rangle\langle\BB\rangle\right) \frac{\partial}{\partial r}\text{ln} \bar{\rho}}, \rule[-7pt]{0pt}{20pt}\\
P_{\rm{FC}} & = & \displaystyle{\left(\langle v'_r\BB'\rangle\right) \frac{\partial}{\partial r}\text{ln} \bar{\rho}}, \rule[-7pt]{0pt}{20pt}\\
P_{\rm{MD}} & = & \displaystyle{-\nabla \times \left(\eta \nabla \times \langle \BB \rangle\right)}. \rule[-7pt]{0pt}{20pt}\\
\end{matrix}
\end{equation}

The expression can be directly used to study the importance of the different terms in the generation of the toroidal magnetic field. 
The time integral of this equation for the longitudinal component gives 
\begin{equation}
\Delta \langle B_{\varphi} \rangle = \int_{t_1}^{t_2} dt \left( P_{\rm{MS}} + P_{\rm{FS}} + P_{\rm{MA}} + P_{\rm{FA}} + P_{\rm{MC}} + P_{\rm{FC}} + P_{\rm{MD}} \right) |_{\varphi}
\end{equation}

\begin{figure}[t!]
  \begin{center}
    \includegraphics[scale=0.55]{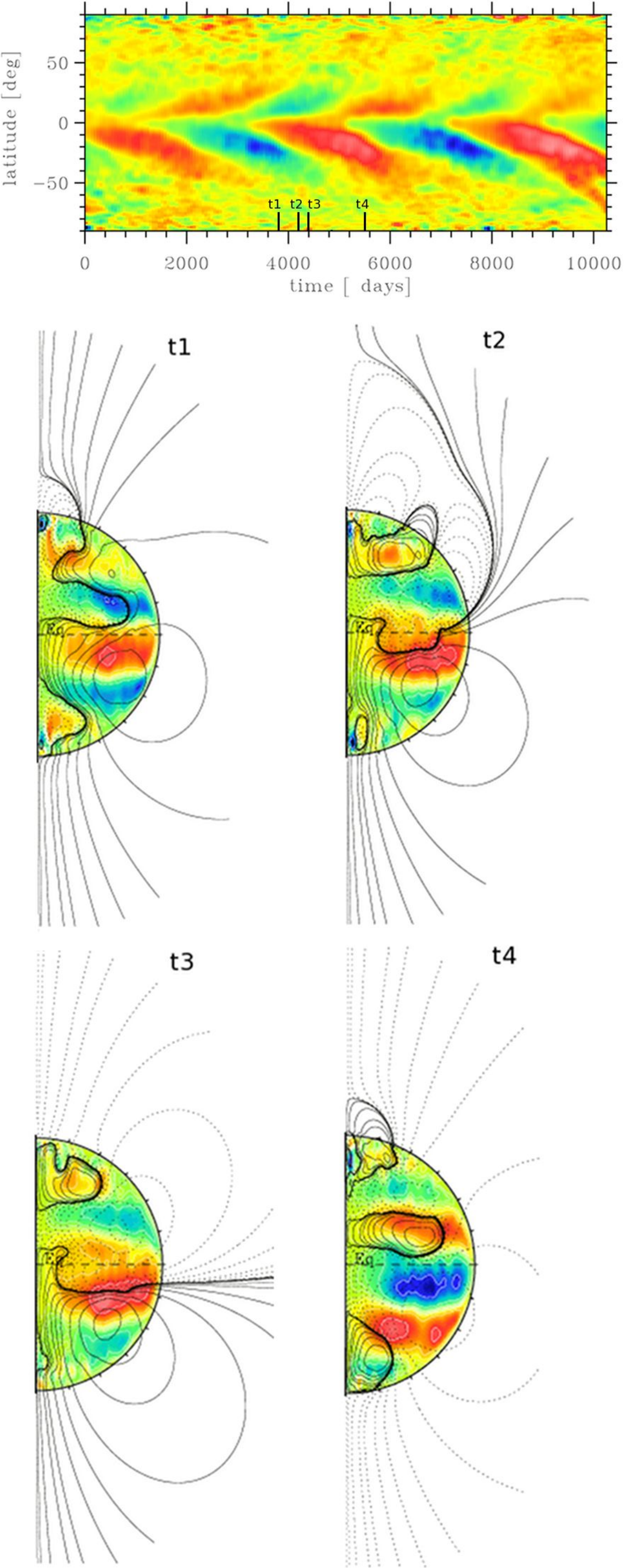}
    \caption{\label{ButtFC}Time evolution of the toroidal magnetic field at the surface of the fully convective model. 
      Cycling period is almost 11 years so a magnetic energy cycle of 5.5 years. 
      Azimuthal averages of both toroidal and poloidal magnetic fields are chosen to show the evolution and reversal of magnetic field. 
      At $t_1$, $B_{\rm{pol}}$ is almost completely positive. 
      Reversal begins at $t_2$ with increasing negative values in $B_{\rm{pol}}$ and decreasing patterns amplitude for $B_{\rm{tor}}$. 
      At $t_3$, negative patterns of $B_{\rm{tor}}$ mostly vanish and $B_{\rm{pol}}$ has opposite signs in the two hemispheres.
      At $t_4$, the reversal is achieved with a fully negative $B_{\rm{pol}}$ and patterns of $B_{\rm{tor}}$ have the opposite signs of those observed at $t_1$.
      The toroidal field in shown in color, with red for positive values and blue for negative ones. 
      Contours of poloidal field are over plotted with solid line for positive values and dashed line for negative ones. 
      This poloidal field is extrapolated outside the star using
      the PFSS model which neglects electrical currents in the corona.
      The lower boundary is given by our MHD simulations and the upper boundary, modeling the effect on the field of the outflowing solar wind, is characterized by an electric current source surface where the field lines are forced to be radial. 
      In this extrapolation we used a source surface radius of $r_{ss} \simeq 2.5 R_*$ (see \cite{Schrijver2003}). 
}
  \end{center}
\end{figure}

In these models, over the seven physical processes that contribute to the toroidal mean field generation 
the mean compression has a negligible role and the mean advection and the fluctuating compression have small contributions. 
Thus in the following analysis we will neglect the mean advection and the mean compression terms since they are negligible in all our simulations. 
The result of this calculation is shown in Figure \ref{MagFieldProd} for the fully convective model, where $t_1$ and $t_4$ are taken at the maximum and minimum of the dipole
component of the magnetic field (see Figure \ref{dipole}). 
Therefore, the interval $\left[t_1;t_4\right]$ captures one magnetic cycle and one magnetic polarity reversal (see Figure \ref{ButtFC}).
In this figure, we notice that $B_{\varphi}$ is mostly created by the mean shear and destroyed by the mean diffusion.

To have a better understanding of the generation of the toroidal field and to compare the different models, these generation terms are integrated over radius and latitude and we look at the toroidal mean energy:
\begin{equation}
\begin{matrix}
\displaystyle{\frac{d\text{TME}}{dt}} & = & \displaystyle{\int_V dV~ \frac{\partial}{\partial t} \frac{\langle B_{\varphi}^2\rangle }{8\pi}}\\
\\
& = & \displaystyle{\int_V dV~\frac{\langle B_{\varphi}\rangle }{4\pi} \cdot } & \displaystyle{\left(P_{\rm{MS}} + P_{\rm{FS}} + P_{\rm{MA}} + P_{\rm{FA}}\right.} \\ 
& & &\displaystyle{\left.+ P_{\rm{MC}} + P_{\rm{FC}} + P_{\rm{MD}} \right)|_{\varphi}}. 
\end{matrix}
\end{equation}
The balance of time-averaged generation terms for the toroidal energy is shown in Figure \ref{bar} for all cases. 
We notice that the toroidal magnetic energy is sustained by the fluctuating compression term and by the action of both mean and fluctuating shear. 
It is annihilate by the mean diffusion and fluctuating advection terms. 
The mean advection and mean compression terms are negligible compared to the other contributions. 
The mean diffusion has the main contribution to the destruction of the toroidal field. 
The fluctuating advection also counteracts the generation of ME, with a smaller contribution. 
The generation of toroidal field can be split in two quasi-constant contributions : the fluctuating compression (FC) and the shear (FS + MS). 
As the star ages, the contribution of the mean shear increases while the one of the fluctuating shear decreases, except for the last simulation where it is almost the same. 
This tells us about the nature of the mean field generation, shear remains critical in these models. 

\begin{figure*}
  \begin{center}
    \includegraphics[scale=0.6]{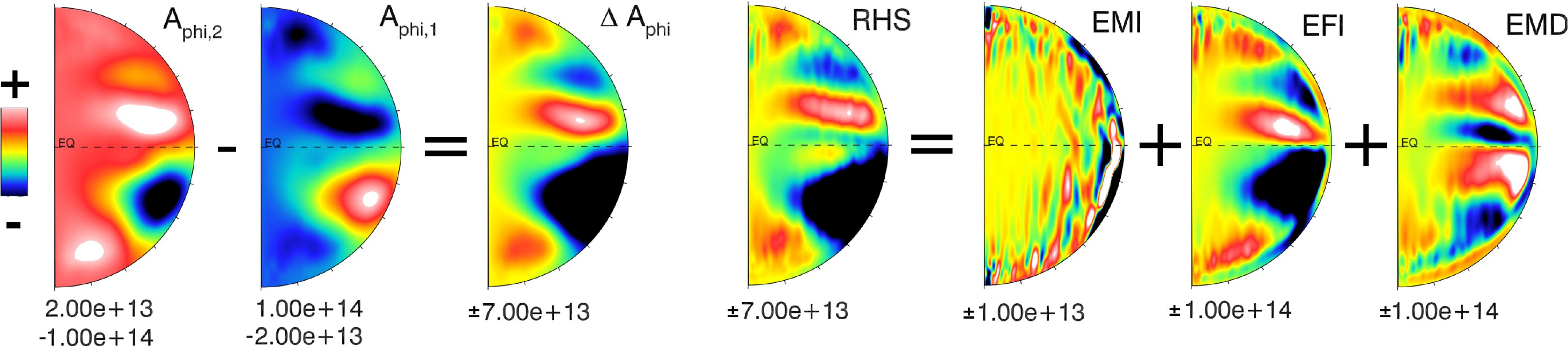}
    \caption{\label{MagFieldProdA}Time evolution the vector potential $A_{\varphi}$ through one magnetic cycle of the fully convective model. 
The three plot on the left show the vector potential at two different times $t_1$ and $t_4$ (see Figure \ref{ButtFC}), and their difference $\Delta A_{\varphi} = \langle A_{\varphi,2} \rangle - \langle A_{\varphi,1} \rangle$. 
The mid plot stands for the sum of the right-hand side. 
The components of the sum are shown individually with the mean EMF, the fluctuating EMF and the resistive diffusion. 
e left 2D plot shows the variation of $B_{\varphi}$ : $\Delta A_{\varphi}  = \left(A_{\varphi,2}-A_{\varphi,1}\right)/(t_4-t_1)$ in G.cm/s. 
}
  \end{center}
\end{figure*}

The production of mean poloidal field is simpler to understand if we represent $\langle \BB_P \rangle$ by its vector potential $\langle A_{\varphi} \rangle$:
\begin{equation}
\langle \BB_{P} \rangle = \nabla \times \langle A_{\varphi} \rangle \hat{\varphi}.
\end{equation}
By uncurling the induction equation \ref{eq::induction} once, we obtain the evolution of the potential vector: 
\begin{equation}
\label{eq::polgeneration}
\frac{\partial \langle \AAA_{\varphi} \rangle}{\partial t} = \left. \langle \mathbf{v} \times \BB \rangle \right|_{\varphi} - \left. \eta \nabla \times \langle \BB \rangle \right|_{\varphi}.
\end{equation}
The first term of the equation is the electromotive force (EMF) coming from the coupling of internal flows and magnetic fields and the second is the ohmic diffusion. 
These terms can also be decomposed into mean and fluctuating components:
\begin{equation}
\begin{matrix}
\displaystyle{E_{\rm{MI}}} & = & \displaystyle{\langle v_r \rangle \langle B_{\theta} \rangle - \langle v_{\theta} \rangle \langle B_r \rangle}\\
\\
\displaystyle{E_{\rm{FI}}} & = & \displaystyle{\langle v'_r B'_{\theta} \rangle - \langle v'_{\theta} B'_r \rangle}\\
\\
\displaystyle{E_{\rm{MD}}} & = & \displaystyle{ - \eta \frac{1}{r} \left(\frac{\partial}{\partial r} \left(r \langle B_{\theta} \rangle \right) - \frac{\partial \langle B_r \rangle}{\partial \theta}\right)}.\\
\end{matrix}
\end{equation} 
This equation is illustrated for the fully convective case by Figure \ref{MagFieldProdA}. 
The main contribution to the generation of poloidal field is the fluctuating EMF 
and it will be studied in detail in the following section through the analysis of the $\alpha-\Omega$ effect. 

\subsection{Assessing the relative contribution of $\alpha-\Omega$ dynamo effects}

The generation of poloidal magnetic field is dominated by the action of the fluctuating EMF: $E_{\rm{FI}} = \mathcal{E}' = \langle \mathbf{v}' \times \BB'\rangle$.
This process can also be interpreted through the $\alpha$-effect approximation which is a first order expansion of $\mathcal{E}'$ around the mean magnetic field and its gradient:
\begin{equation}
\langle \mathcal{E}' \rangle_i = \alpha_{ij} \langle \BB \rangle_j + \beta_{ijk} \nabla \langle \BB\rangle + \mathcal{O} \left( \partial \langle \BB \rangle /\partial t, \nabla^2 \langle \BB \rangle \right)
\end{equation}
with $\alpha_{ij}$ a rank-two pseudo-vector and $\beta_{ijk}$ a rank-three tensor. In the following, we will neglect the $\beta$ term. 
However, this will increase the systematic error when estimating the $\alpha$ term. 
Thus a single-value decomposition (SVD) including the $\beta$-effect has been calculated in order to provide a lower-bound on the systematic error \citep[][]{Augustson2015}. 
In the following analysis, $\alpha$ has been decomposed into its symmetric and antisymmetric components 
\begin{equation}
\label{eq::alphagamma}
\alpha \langle \BB\rangle = \alpha_S \langle \BB \rangle + \gamma \times \langle \BB \rangle 
\end{equation}
with
\begin{equation}
\alpha_S=
\left[
\begin{matrix}
\alpha_{(rr)} & \alpha_{(r\theta)} & \alpha_{(r\varphi)} \\
\alpha_{(r\theta)} & \alpha_{(\theta\theta)} & \alpha_{(\theta\varphi)} \\
\alpha_{(r\varphi)} & \alpha_{(\theta\varphi)} & \alpha_{(\varphi\varphi)} \\
\end{matrix}
\right]
\quad\hbox{ and }\quad \gamma=
\left[
\begin{matrix}
\gamma_{r} \\
\gamma_{\theta} \\
\gamma_{\varphi} \\
\end{matrix}
\right].
\end{equation}

In our study, we will focus on the efficiency of the $\alpha$-effect and on the characterization of our dynamo through the relative influence of its regenerating terms. 
At first, one interesting measure of these dynamo is to quantify the capacity of the convective flows to regenerate mean magnetic fields. 
This can be evaluated by finding the average magnitude of an estimated $\alpha$-effect relative to the rms value of the non axisymmetric velocity field 
\begin{equation} \label{eq::alphaefficiency}
E \simeq \left\langle \frac{\alpha}{\text{v}_{\rm{rms}}} \right\rangle = \frac{3}{2(r_2^3-r_1^3)} \times \sum_{i,j} \iint dr d\theta r^2 \sin \theta \sqrt{\frac{\alpha_{ij} \alpha^{ij}}{\lbrace \mathbf{v}'\cdot \mathbf{v}'\rbrace}}
\end{equation}
where $\lbrace \mathbf{v}'\cdot \mathbf{v}'\rbrace$ is the sum of the diagonal elements of the Reynolds stress tensor averaged over time and over all longitudes. 
If we want to refine the analysis, we can use the equation \ref{eq::alphaefficiency} to provide a measure of the importance of each component of $\alpha$ as
\begin{equation}
\label{eq::alphacomponentefficiency}
\begin{matrix}
\displaystyle{\varepsilon_{ij}} & = & \displaystyle{\frac{E_{ij}}{E}} & & \\
&\simeq & \displaystyle{\frac{1}{E} \left\langle\frac{\alpha_{ij}}{\text{v}_{\rm{rms}}} \right\rangle} & = & \displaystyle{\frac{3}{2E(r_2^3-r_1^3)} \iint dr d\theta r^2 \sin \theta \sqrt{ \frac{\alpha_{ij} \alpha^{ij}}{\lbrace \mathbf{v}' \cdot \mathbf{v}'\rbrace}}} \\
& & \\
& & & = & \left[\begin{matrix} \varepsilon_{(rr)} & \varepsilon_{(r\theta)} & \varepsilon_{(r\varphi)} \\\varepsilon_{\gamma_{\varphi}} & \varepsilon_{(\theta\theta)} & \varepsilon_{(\theta\varphi)} \\ \varepsilon_{\gamma_{\theta}} & \varepsilon_{\gamma_r} & \varepsilon_{(\varphi\varphi)} \end{matrix} \right]
\end{matrix}
\end{equation}
with $\varepsilon_{(xx)} = \displaystyle{\frac{\alpha_{(xx)}}{E}}$ and $E_{\gamma x} = \displaystyle{\frac{\gamma_x}{E}}$. 
By calculating this matrix, see Table \ref{alphaomega}, we notice that for the antisymmetric part $\gamma$, the predominant term is always $\gamma_{\varphi}$ that impacts the poloidal component of the magnetic field. 
$\gamma_r$ and $\gamma_{\theta}$ have the same order of magnitude and are between 3 and 18 times smaller than $\gamma_{\varphi}$. 
By looking at the symmetric part $\alpha_{\rm{S}}$, we see the same trend. 
The predominant term is either $\alpha_{(rr)}$ or $\alpha_{(\theta\theta)}$ which both act on the poloidal component of the magnetic field and 
the smallest term is always $\alpha_{(\varphi\varphi)}$ which is at least two times smaller than the predominant term. 
Thus we may conclude that the relative influence of the poloidal field regeneration by the EMF is more important than the toroidal field regeneration. 
We can quantify this relative influence $\alpha_{\rm{P}}/\alpha_{\varphi}$: 
\begin{equation}
\frac{\alpha_{\rm{P}}}{\alpha_{\varphi}} = \frac{3}{2(r_2^3-r_1^3)} \times \iint dr d\theta r^2 \sin \theta \left| \frac{\langle \BB_{\rm{P}} \rangle \cdot \nabla \times \langle \mathcal{E}' \rangle}{\langle B_{\varphi} \hat{\varphi} \cdot \nabla \times \langle \mathcal{E}'\rangle}\right|.
\end{equation}

\begin{table}[!th]
\begin{center}
\caption{$\alpha-\Omega$ effect}\label{alphaomega}
\vspace{0.2cm}
\begin{tabular}{|c|ccc|c|c|}
\cline{2-6}
\multicolumn{1}{c}{}  & \multicolumn{3}{|c|}{$\alpha$ tensor} & $\Omega/\alpha_{\varphi}$ & $\alpha_{P}/\alpha_{\varphi}$ \rule[-9pt]{0pt}{24pt}\\ \hline
                      & 0.104                        & 0.114                        & 0.130 &       &       \rule[-7pt]{0pt}{20pt} \\
FullConv              & \cellcolor{light-gray} 0.114 & 0.127                        & 0.147 & 4.20  & 12.5  \rule[-7pt]{0pt}{20pt}\\
                      & \cellcolor{light-gray} 0.082 & \cellcolor{light-gray} 0.086 & 0.095 &       &       \rule[-7pt]{0pt}{20pt}\\ \hline
                      & 0.133                        & 0.129                        & 0.115 &       &       \rule[-7pt]{0pt}{20pt}\\
$20\%$                & \cellcolor{light-gray} 0.120 & 0.116                        & 0.124 & 2.08  & 10.8  \rule[-7pt]{0pt}{20pt}\\
                      & \cellcolor{light-gray} 0.084 & \cellcolor{light-gray} 0.087 & 0.073 &       &       \rule[-7pt]{0pt}{20pt}\\ \hline
                      & 0.203                        & 0.126                        & 0.103 &       &       \rule[-7pt]{0pt}{20pt}\\
$40\%$                & \cellcolor{light-gray} 0.149 & 0.097                        & 0.098 & 3.42  & 19.1  \rule[-7pt]{0pt}{20pt}\\ 
                      & \cellcolor{light-gray} 0.085 & \cellcolor{light-gray} 0.070 & 0.069 &       &       \rule[-7pt]{0pt}{20pt}\\ \hline
                      & 0.279                        & 0.137                        & 0.061 &       &       \rule[-7pt]{0pt}{20pt}\\ 
$60\%$                & \cellcolor{light-gray} 0.194 & 0.102                        & 0.056 & 4.39  & 21.2  \rule[-7pt]{0pt}{20pt}\\ 
                      & \cellcolor{light-gray} 0.077 & \cellcolor{light-gray} 0.053 & 0.040 &       &       \rule[-7pt]{0pt}{20pt}\\ \hline
                      & 0.188                        & 0.155                        & 0.032 &       &       \rule[-7pt]{0pt}{20pt}\\ 
$70\%$                & \cellcolor{light-gray} 0.099 & 0.084                        & 0.024 & 1.54  & 9.19  \rule[-7pt]{0pt}{20pt}\\ 
                      & \cellcolor{light-gray} 0.211 & \cellcolor{light-gray} 0.171 & 0.035 &       &       \rule[-7pt]{0pt}{20pt}\\ \hline
\end{tabular}
\end{center}
\textbf{Note:} Results of the dynamo analysis on our PMS models. 
The first column represents the $\alpha$ tensor with its symmetric: $\alpha_{\rm{s}}$ (white background) and antisymmetric: $\gamma$ (gray background) portions (see Eq \ref{eq::alphagamma}). 
The middle column gives the relative importance of the $\Omega$-effect to the $\alpha$-effect for the regeneration of the toroidal field. 
The last column quantifies the ratio of the $\alpha$-effect used for the regeneration of the poloidal magnetic field to the one used for the regeneration of the toroidal field.
\end{table}
With Table \ref{alphaomega}, we confirm the predominance of the poloidal field regeneration over the toroidal field regeneration for all our models. 
This quantity also enables us to see the evolution of this relative influence over all our models. 
The influence of the poloidal part of $\alpha$ in the 20\% model is smaller than in the fully convective model. 
This decrease cannot be due to the rotation rate since the two models have the same stellar rotation rate. 
The main difference between these models is the size and geometry of the convective envelope. 
In the following models, with a decrease in size of the convective envelope and an increase of the rotation rate, a trend appears with an increase of the weight of the poloidal part in $\alpha$ as the star evolves along the PMS. 
The 70\% model behaves differently because of a change of trend in $L_*$ that impacts $\rm{v}_{\rm{rms}}$. 

As seen in equation \ref{eq::torgeneration}, the $P_{MS}$ term plays an important contribution for the generation of toroidal magnetic field. 
This term, called the $\Omega$-effect, represents the action of mean shearing, i.e. differential rotation, on the poloidal magnetic field: 
\begin{equation}
P_{MS} = r \sin \theta ~\langle \BB_P \rangle \cdot \nabla \Omega.
\end{equation}

In mean-field theory, the regeneration of the toroidal field can both be due to the $\alpha$-effect, coming from the fluctuating EMF, and to the $\Omega$ effect that acts on the poloidal field through differential rotation. 
In all our models, we note that the regeneration of $\langle B_{\varphi} \rangle$ by the $\alpha$-effect is small, compared to the one of $\BB_{\rm{pol}}$.
Therefore, we now want to measure the relative influence of the $\Omega$-effect to that of the $\alpha$-effect, since the toroidal magnetic field can be regenerated through both effects:
\begin{equation}
\frac{\Omega}{\alpha_{\varphi}} = \frac{3}{2(r_2^3-r_1^3)} \times \iint dr d\theta r^2 \sin \theta \left| \frac{r \sin \theta \langle B_{\varphi} \rangle \langle \BB_{P}\rangle \cdot \nabla \langle \Omega \rangle}{\langle B_{\varphi}\rangle \hat{\varphi} \cdot \nabla \times \langle \mathcal{E}'\rangle}\right|.
\end{equation}
As the rotation rate remains constant and the convective envelope size decreases, the influence of the $\Omega$-effect decreases with respect to the $\alpha$-effect. 
In the following models, which have a growing radiative core and an increasing rotation rate, the influence of $\Omega$-effect increases with respect to $\alpha_{\varphi}$.
The poloidal $\alpha$ effect also becomes more and more predominant with respect to $\alpha_{\varphi}$.
This can be understood as the $\Omega$-effect is strongly link to the differential rotation and as we see in section \ref{KEMHD} that the contrast in differential rotation grows as the star evolves along the PMS. 
The 70\% model behaves differently. 
Stellar luminosity increases in this model, whereas it decreases in the others. 
As luminosity increases, we increases both viscous and magnetic diffusivities to keep consistent Reynolds numbers. 
This increasing magnetic diffusivity possibly explains the behavior of the 70\% model.

\subsection{Time evolution and magnetic cycles}\label{cycles}

Time evolution of the magnetic field through the PMS, as shown in the time-latitude plot in Figure \ref{fig:Butterfly}, 
clearly possesses a cyclic behavior for three out of five cases.
For illustrative purposes, we will now discuss one of these cyclic dynamo cases, namely the fully convective one.
In Figure \ref{ButtFC}, we display a zoomed in version of its time-latitude diagram along with meridional cut for specific times samples.
We notice that we have a dynamo cycle of almost 11 years, hence commensurable with what we know about the typical length of magnetic cycles in solar-like stars. 
We also notice a beginning of reversal at the end of the evolution of the second model, the one with a 20\% radiative zone and we decide to continue the computation of this model and we see that a cycling dynamo appears. 

\begin{figure}
  \begin{center}
    \includegraphics[scale=0.5]{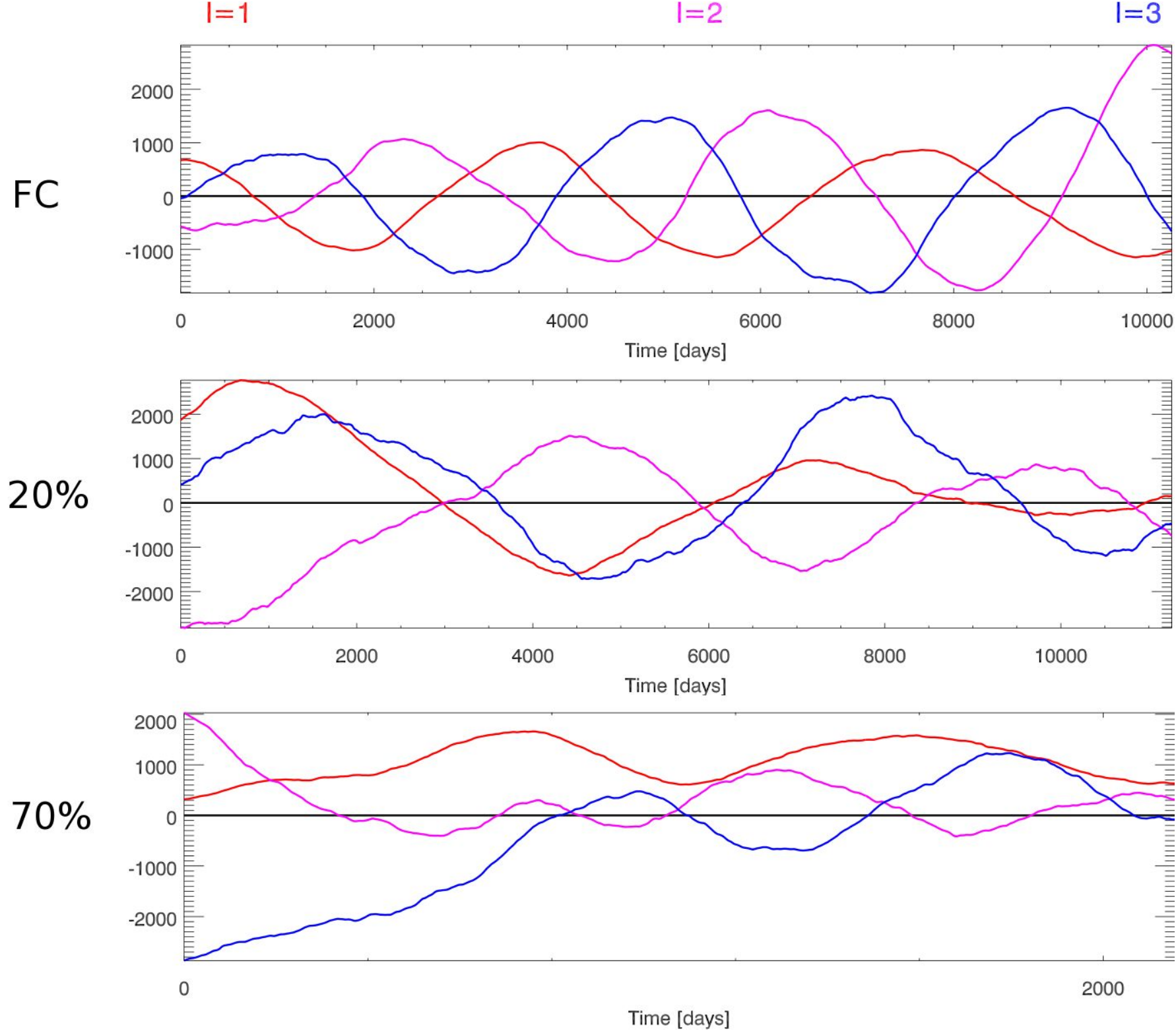}
    \caption{\label{dipole}The amplitudes $\mathcal{D}$, $\mathcal{Q}$ and $\mathcal{O}$ are shown for the cycling models : FC, 20\% and 70\%. 
      We note an increase of the cycle period length. 
      No clear phasing relationship between the three components can be established, except for the 20\% case that show that symmetric and anti-symmetric dynamo mode are in opposition of phase. 
      We also notice that in the 70\% model there is no dipole reversal, even if it presents a cycling behavior.
}
  \end{center}
\end{figure}

The temporal evolution of the dipole, quadrupole and octupole moments of the magnetic field can by measured through the amplitudes of the $m=0$ mode of the radial component of the magnetic field $B_r$. 
These amplitudes are calculated by the expressions 
\begin{equation}
\begin{matrix}
\mathcal{D} & = & \displaystyle{\sqrt{\frac{3}{4\pi}} \int B_r \cos \theta \sin \theta d\theta d\phi} ,\\
\\
\mathcal{Q} & = & \displaystyle{\frac{1}{2} \sqrt{\frac{5}{4\pi}} \int B_r \left(3\cos^2 \theta -1 \right) \sin \theta d\theta d\phi} ,\\
\\
\mathcal{O} & = & \displaystyle{\frac{1}{2} \sqrt{\frac{7}{4\pi}} \int B_r \left(5\cos^3 \theta -3\cos \theta \right) \sin \theta d\theta d\phi} ,
\end{matrix}
\end{equation}
with the integral solid angle is at a fixed radius. 
The amplitudes $\mathcal{D}$, $\mathcal{Q}$ and $\mathcal{O}$ are shown for both models in Figure \ref{dipole}. 
The measurements are done at $96\%$ of $r_{\rm{top}}$.

As seen above, our simulations show equatorward propagation of the magnetic field (see Figure \ref{fig:Butterfly}). 
In three of those models, FC, 20\% and 70\%, we even see dynamo cycles. 
The equatorward propagation in kinematic $\alpha-\Omega$ dynamo is generally attributed to the propagation of a dynamo wave. 
In this theory, the propagation direction of the dynamo wave is given by the Parker-Yoshimura rule \citep[][]{Parker1955,Yoshimura1975}
\begin{equation}\label{Stheta}
\mathbf{S} = - r \sin \theta \bar{\alpha} \mathbf{\hat{e}_{\varphi}} \times \nabla \frac{\Omega}{\Omega_0}
\end{equation}
with $\bar{\alpha} = -\tau_0 \langle \mathbf{v} \cdot \boldsymbol{\varpi}\rangle/3$, $\tau_0$ is the convective overturning time and $\boldsymbol{\varpi} = \nabla \times \mathbf{v}$ the vorticity and $\langle \cdot\rangle$ is the azimuthal average. 
\begin{figure}
  \begin{center}
    \includegraphics[scale=0.37]{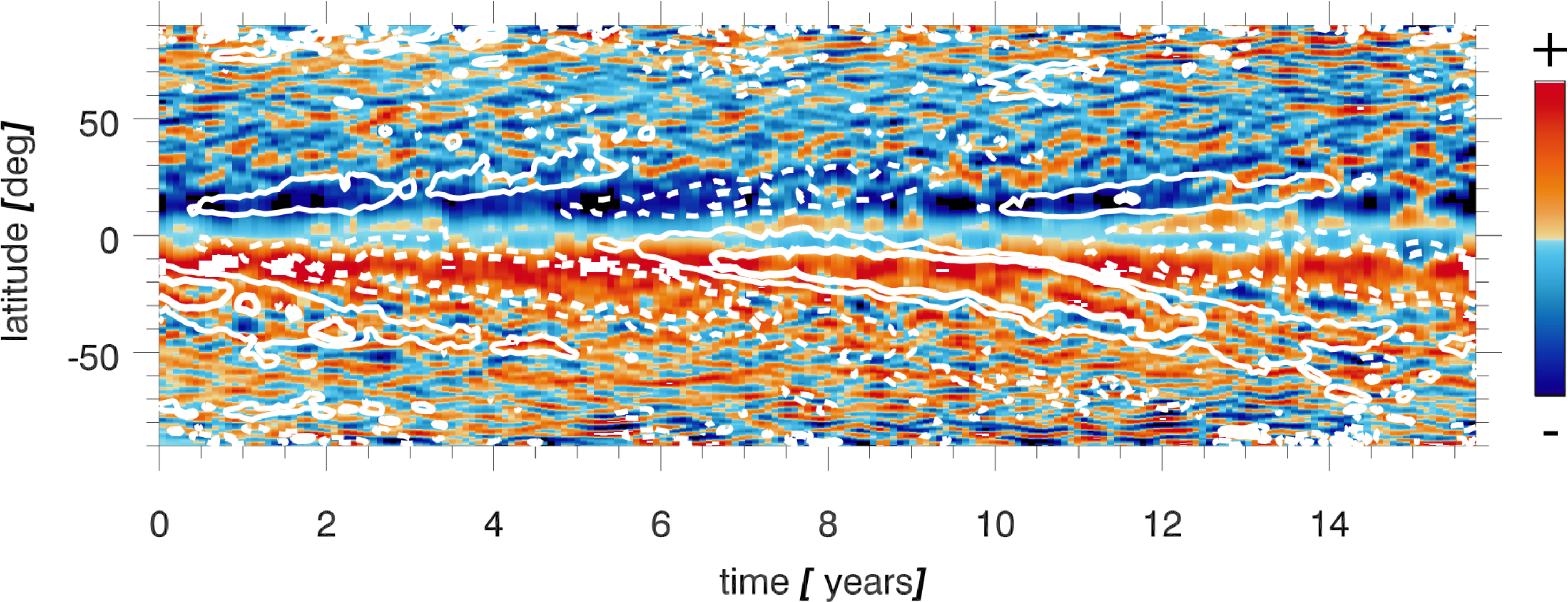}
    \caption{\label{PY}Latitudinal component of the propagation direction of a dynamo wave $S_{\theta}$ (see equation \ref{Stheta}), for the fully convective model. 
      Red contours denote positive southward direction.
      Overplotted in white are isocontours of $\langle B_{\varphi}\rangle$ at 1 and 2 kG, with solid contours being of positive polarity and dashed of negative polarity. 
We note the good agreement of the Parker-Yoshimura rule with dynamo branch shown in the butterfly diagram of Figure \ref{ButtFC}.
}
  \end{center}
\end{figure}
To further illustrate the nature of dynamo action realized in our simulations, we calculated $S_{\theta}$ for one of the models with a cycle dynamo : the fully convective one. 
In Figure \ref{PY}, the Parker-Yoshimura rule is respected as $S_{\theta} <0$ (resp. $S_{\theta} > 0$) in the northern (resp. southern) hemisphere 
implies that the dynamo wave propagate from the equator to the poles at low latitudes, as in the butterfly diagram displayed in Figure \ref{ButtFC} for the same fully convective case. 

\subsection{Magnetic field evolution in the radiative zone}

\begin{table*}
\begin{center}
\caption{Magnetic energy decomposition}\label{MenergiesRZ} 
\vspace{0.2cm}
\begin{tabular}{cccccccc}
\tableline
\tableline
\\ [-1.5ex]
 Case & EM$^{RZ}$       & ME$^{RZ}$              & ME$_{\rm{pol}}$$^{RZ}$ & ME$_{\rm{tor}}$$^{RZ}$ & ME$^{CZ}$              & ME$_{\rm{pol}}$$^{CZ}$ & ME$_{\rm{tor}}$$^{CZ}$ \\ [0.5ex]
      & ($10^{32}$ erg) & ($10^4$ erg.cm$^{-3}$) & ($10^4$ erg.cm$^{-3}$) & ($10^4$ erg.cm$^{-3}$) & ($10^5$ erg.cm$^{-3}$) & ($10^5$ erg.cm$^{-3}$) & ($10^5$ erg.cm$^{-3}$) \\ [0.8ex]
\tableline
\tableline
\\[-1.5ex]
      &                 &                        &                        &                        &                        &                        &                        \\ [-1.5ex]           
FC    & --              & --                     &   --                   & --                     & 5.62                   & 3.05 (54.3\%)          & 2.57 (45.7\%)          \\
20 \% & 2.50            & 3.40                   & 1.87 (55.15\%)         & 1.52 (44.86\%)         & 7.49                   & 3.59 (47.8\%)          & 3.91 (52.2\%)          \\
40 \% & 22.7            & 8.46                   & 2.72 (32.13\%)         & 5.74 (67.87\%)         & 10.7                   & 5.60 (52.1\%)          & 5.14 (47.9\%)          \\
60 \% & 121             & 33.7                   & 19.7 (58.48\%)         & 14.0 (41.52\%)         & 21.0                   & 11.6 (55.5\%)          & 9.29 (44.5\%)          \\
70 \% & 41.3            & 8.6                    & 6.34 (73.5\%)          & 2.29 (26.5\%)          & 8.8                    & 4.68 (53.2\%)          & 4.12 (46.8\%)          \\ [0.5ex]
\tableline
\tableline
\end{tabular}
\end{center}
\normalsize
\textbf{Note:} The first column gives the global magnetic energy (EM) in the radiative zone (in erg). 
The following columns show energy densities with the magnetic energy (ME) divided into two its toroidal and poloidal part (ME$_{\rm{tor}}$, ME$_{\rm{pol}}$). 
All energy densities are averaged over a period of 400 days and reported in erg cm$^{-3}$.
\end{table*}

Observations conducted on the activity of massive stars have shown that some of these stars possess a strong surface magnetic fields \citep[][]{Babcock1947,Mathys2001,Donati2006,Ferrario2005,Beuermann2007,Becker2003,Auriere2007}. 
However these fields are drastically different of those observed in convective stars. 
None of these have very small scales fields, they all presented strong low $l$ components (dipole, quadrupole or octupole). 
Moreover none of these non convective stars have differential rotation which is mostly the case of our radiative cores. 
Hence the lack of ingredients for a dynamo generation leads us to infer, as \cite{Braithwaite2008}, that these magnetic fields in radiative zone are \textit{fossil fields}, i.e. stable fields that evolves on diffusive time scales. 
These fields are very sensitive to instabilities and many of them are unstable and disappears quickly. 
This is coherent with observations since many massive stars do not seem to possess surface magnetic field. 
These instabilities were studied by \cite{Tayler1973} and \cite{MarkeyTayler1973} which showed that both purely poloidal and toroidal magnetic fields cannot be stable : we need a mixed configuration poloidal-toroidal. 
\cite{Braithwaite2009} gave a quantitative upper limit to that stable mixed configuration : the poloidal magnetic energy must be less that 80\% of the total magnetic energy. 

\begin{figure}
  \begin{center}
    \includegraphics[scale=0.4]{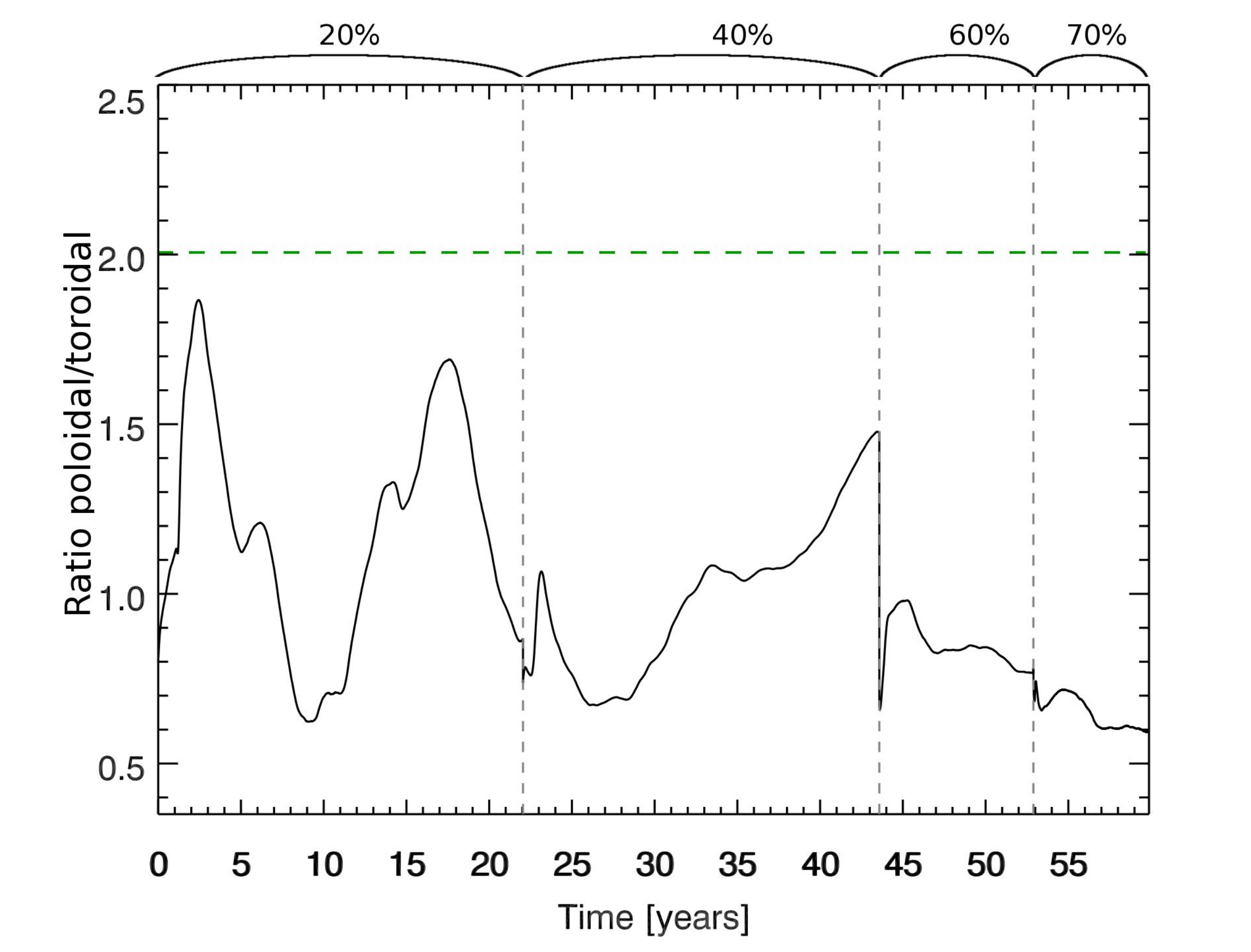}
    \caption{\label{ratiostab} Evolution of the ratio $B_{\rm{pol}}/B_{\rm{tor}}$ over the PMS. 
}
  \end{center}
\end{figure}

Thus we use the decomposition of the magnetic field: 
\begin{equation}
\text{ME} = \text{ME}_{\rm{pol}} + \text{ME}_{\rm{tor}}.
\end{equation} 
This decomposition can be analyzed in both convective and radiative zones (as shown in Table \ref{MenergiesRZ}). 
These values are taken averaged over the last 400 days of each simulations. 
Hence each final magnetic field fulfill the Braithwaite criteria: ME$_{\rm{pol}}$$^{RZ}< 0.8$ME$^{RZ}$. 
We also notice that the magnetic energy is mostly in equipartition between the poloidal and the toroidal parts in the radiative core, but also in the convective envelop. 
Hence, each time that the radiative core grows from a radius $r_{bcz,n}$ to $r_{bcz,n+1}$, the magnetic field ``introduced'' by this change in size already fulfill the stability criteria. 
Since the relaxation does not change much the partition of $\BB$ between poloidal and toroidal, the magnetic field logically fulfill the stability criteria. 
By looking at the table of magnetic energies (table \ref{MenergiesRZ}), we notice that this limit is actually never reaches (when our models are relaxed). 
We decided to track the ratio $B_{\rm{pol}}/B_{\rm{tor}}$ all along our simulations to see if this stability criteria was always fulfilled. 
Since $ME_{\rm{pol}}/ME$ must be lower than 0.8, we are looking if $B_{\rm{pol}}/B_{\rm{tor}} < 2$. 
Figure \ref{ratiostab} shows the evolution of that ratio along the PMS. 
Hence we notice that even during the transient time between two different internal structures, the stability criteria defined by Braithwaite is fulfilled. 
As we propagate the magnetic field from one stellar structure to the following one, 
we need to check that we compute the MHD simulation long enough to enable $\BB$ to relax in the radiative core. 
Relaxation takes place on several Alfvén times $t_a = v_a/D_{RZ}$ where $D_{RZ}$ is a characteristic length scale of the relaxation in the radiative zone. 
By looking at the internal structure of the model $n+1$, we have three different areas. 
$r > r_{bcz,n+1}$ defines the convective envelop of the model. 
$r < r_{bcz,n+1}$ is the radiative core of the simulation. 
This zone can be split in two : $r < r_{bcz,n}$ that was already radiative in the previous model 
and $r_{bcz,n} < r < r_{bcz,n+1}$ that was convective in the previous model and becomes radiative in this model. 
In the first area, the relaxation process occurred in previous models as it was already radiative. 
The relaxation process that we look at in the $n+1$ model thus occurs in the portion limited by the radii : $r_{bcz,n}$ and $r_{bcz,n+1}$. 
That is why we choose to define $D_n$ as $D_n = r_{bcz,n+1}-r_{bcz,n}$ rather than as the radius of the radiative zone of the model. 
\begin{table}
\begin{center}
\caption{Alfvén time for the radiative zone}\label{alfven}
\vspace*{0.2cm}
\begin{tabular}{cccccc}
\tableline
\tableline
\\ [-1.5ex]
Case     & $v_a$         & $D_n$        & $t_{a}$ & \multicolumn{2}{c}{$t_{\rm{model}}$} \\ [0.5ex]
         & (cm.s$^{-1}$) & $10^{10}$ cm & (days)  & (days) & ($t_a$) \\ [0.8ex]
\tableline
\tableline
         &               &              &         &        &         \\ [-1.5ex]
20 \%    & 181           & 2.6          & 1660    & 8062   & 4.87    \\
40 \%    & 194           & 2            & 1190    & 7875   & 6.61    \\
60 \%    & 202           & 1.47         & 838     & 2454   & 2.93    \\
70 \%    & 461           & 0.69         & 1320    & 2555   & 1.94    \\ [0.5ex]
\tableline
\tableline
\end{tabular}
\end{center}
\textbf{Note:} Relaxation time in the radiative zone. 
The first column gives the Alfvén speed in the core of our simulations. 
The second one shows the characteristic length scale of the relaxation. 
The Alfvén time is defined by $t_a = v_a/D_{RZ}$.
The last columns shows the computational time of the MHD models in days and in Alfvén time. 
\end{table}
The values of Alfvén times for our simulations are given in Table \ref{alfven}. 
We see that each MHD model is evolved over few Alfvén times to insure the relaxation of the magnetic field in the radiative zone.
We conclude that interestingly dynamo action tends to generate mixed fields whose properties satisfy stability criteria in stratified radiative core. 
This result on the stability of the fossil field left over by dynamo action as the convective envelope becomes shallower along the PMS is a direct outcome of our set of simulations and could not be easily anticipated. 
It has direct consequences on the geometry of fossil field that can be expected in solar-like star's radiative core as stable mixed poloidal-toroidal configurations should be favored.

\section{Discussion and conclusion}\label{Conclusion}

During the PMS, between the protostellar phase and the ZAMS,  the stellar radius decreases due to gravitational contraction. 
As the star contracts, internal temperature and pressure increase, opacity drops in the stellar core and a radiative zone appears and grows within the star. 
Moreover, the stellar contraction causes an increase of the rotation rate due to angular momentum conservation. 
Hence we expect the star's dynamical properties to vary significantly and these variations are key to characterize. 
In order to do so, we have developed a series of 3D MHD simulations of stellar convective dynamo. 
To make this study more realistic, we used to setup the spherically symmetric background state of our 3D models, radial profiles obtained from 1D stellar evolution model at various stages along the PMS evolutionary track. 
We choose five different models that represent the star at specific ages of the PMS with different rotation rates and radiative radii. 
At first, we run hydrodynamical 3D simulations of these models in order to equilibrate internal flows and coupling between the radiative core and the convective envelop. 
Then we inject a magnetic field into the fully convective model and once the MHD simulation is equilibrated we inject the resulting magnetic field into the following model. 
We compute all the MHD simulations by propagating the magnetic field into all HD models. 

Our five MHD simulations show the mutual influence of the internal magnetic field and internal flows as the star evolves along the PMS. 
As seen in section \ref{DynamoAction}, the introduction of the magnetic field in the hydrodynamical models leads to important modifications of the internal mean flows and the convective patterns. 
As the differential rotation profiles are quenched by the influence of the Maxwell stresses, the radial convective patterns are larger, since they are less sheared. 
The internal magnetic field also has an notable impact on the angular momentum transport since there are two additional contributions : the Maxwell stresses and the large scale magnetic torques. 
Indeed, in hydrodynamical models, the inward propagation of the angular momentum is due to viscous diffusion whereas in the MHD simulation this contribution becomes small, given the weak differential rotation present in these MHD simulations, and the inward propagation is sustained by the large scale magnetic torques and Maxwell stresses. 

As the star ages along the PMS, we analyze the evolution of the magnetic energy. 
The proportion of mean field energy (TME + PME) decreases strongly from $47\%$ to $6\%$ of the total magnetic energy. 
In all models, the poloidal mean energy prevails on the toroidal mean energy. 
The decrease of the mean energy is coherent with results found by \cite{Gregory2012} in which the magnetic field is less axisymmetric and more complex as the radiative core is bigger. 

As the magnetic field is propagated through the PMS models, we study the evolution of its topology and amplitude as well as its generation through the $\alpha-\Omega$ effect. 
At first, we notice that in both zones, either convective or radiative, the magnetic energy increases as the star ages. 
In our five MHD simulations, we notice that the topology of the magnetic field changes strongly as the star ages. 
Since we follow the evolutionary path of a solar-like star, both rotation rate and aspect ratio of the convective envelop change as the star evolves along the PMS.
The specific influence of each parameter is not always easy to disentangle in our study. 
The influence of internal structure was studied by \cite{Gregory2012}. 
The results of this study show that as the radiative core becomes bigger the dipole components drop and the magnetic field becomes more and more complex. 
These properties are coherent with the results obtained with our MHD simulations 
with the ratio $B_{axi}/B_{non,axi}$ dropping from $0.61$ to $0.18$ and the dipole field strength decreasing from $0.12$ to $0.042$. 
By plotting these two quantities as a function of the Rossby number, we want to analyze the influence of rotation on magnetic topology. 
Rossby numbers of our simulations are quite close to the transition value $R_o=0.1$ found in several studies \citep[][]{Pizzolato2003,Wright2011,Reiners2014,Schrinner2012}. 
We notice that $f_{\rm{dip}}$ is smaller when $R_o < 0.1$. 
For small Rossby numbers, the dipole field strength increases slightly but we need more simulations to know if this value is the upper limit 
or if for lower Rossby number $f_{\rm{dip}}$ continues to grow and shows that $R_o = 0.1$ is a sharp transition.

The generation of the mean magnetic field shows that as the convective zone becomes shallower and the rotation rate increases, the $\Omega$ effect becomes predominant in the generation of the mean toroidal magnetic field. 
Moreover the $\alpha$-effect tends to generate more poloidal field than toroidal one. 
Hence, in each model, we see an $\alpha-\Omega$ dynamo. 
Three out of our five MHD simulations display a magnetic cycle. 
In all cyclic cases, the time latitude diagram of the longitudinally averaged toroidal magnetic field shows a clear poleward branch starting from low latitude (see Figure \ref{ButtFC}). 
This magnetic field propagation is also compatible with the $\alpha-\Omega$ dynamo concept as it satisfies Parker-Yoshimura rule (see Figure \ref{PY}).

As the radiative zone grows in the star, we observe that, in all models, the magnetic field in the core, left over by the convective dynamo, is stable regarding the limit given by \cite{Braithwaite2008}: 
$E_{\rm{pol}}/E_{\rm{tot}} < 0.8$. 
The magnetic field in the radiative core of the star originates from the relaxation of the dynamo field coming from the previous stellar evolution phase in our sequence of models. 
By looking at this dynamo field, we notice, that in all convection zones, the magnetic field that comes from the dynamo action also fulfill the stability criteria $E_{\rm{pol}}/E_{\rm{tot}} < 0.8$ 
even if this has no obvious consequence in that zone. 
It seems that the relaxation of the field preserves this feature of the field and explains that all relaxed fields in radiative core are stable for the stability criteria. 

The global properties of the magnetic fields we obtain in our study also have direct consequences on the coronae of PMS stars. 
We have computed the change of the Alfvén radius (e.g. the radius where the stellar wind decouples from the star) 
that such topological and rotation state implies,
following the prescription described in \cite{Reville2015b}. 
We find that the Alfvén radius shrink from about 33 $R_{\odot}$ to 10 $R_{\odot}$. 
This is in good qualitative agreement with the recent work of \cite{Reville2016} who computed realistic 3D stellar wind along the evolutionnary track 
of a solar mass star using spectro-polarimetric maps from \cite{Folsom2016}. 
We intend in the near future to use the magnetic  field coming out of our dynamo simulations to compute similar 3D wind solutions along the PMS. 
This will allow us to assess the loss of mass and angular momentum, that must also vary significantly given the large change of Alfvén radius we have identified.

The 3D simulations studied in this paper, five HD progenitors and five MHD dynamo simulations, are an idealistic representation of the evolution of solar-like stars along the PMS.
For instance, the turbulence degree in our numerical models is far from reaching that of a real star. 
We try to keep a comparable degree of turbulence in all our models while taking into account computational constraints (each simulation has required on average 1 Mh node hours).
One could also consider a systematic parametric study of each model by varying for instance, in the simulations, their Reynolds and Prandlt numbers to better assess their sensitivity to parameter change. 
Still we are confident that trends presented in this work are robust as we have shown that there are in qualitative agreement with observations and compatible with previous numerical studies of stellar dynamos.
One challenge of this study is its \textit{serial} aspect : before computing the model $n$, we need to reach an equilibrium state, in the statistically stationary sense, for the magnetohydrodynamical model $n-1$. 
Indeed, as already stated, we need the magnetic field of the simulation $n-1$ to initialize the model $n$. 
This serial aspect makes this study complex to compute as each model take several months to settle down. 
One improvement to this analysis would be to increase the number of models and make the gap between the different rotation rates and radiative radii smaller to have a smoother evolution of $\BB$ and trends. 
This may be done by changing our way of simulating the stellar evolution.

It is important to notice that in this study, we choose our models to follow an astrophysical path along the PMS. 
A logical follow-up is therefore to apply this analysis to the evolution of solar-like stars along the following step of stellar evolution, i.e. the main sequence. 
In that study the main parameter will be the decrease of the rotation rate as the star is braked by the solar wind and the internal stellar structure of the star is fixed during this evolutionary phase. 
An additional study would be to study the impact of stellar structure with a fixed rotation rate.
We have started doing such studies and their results will be reported in future communications.

\acknowledgements

First we wish to thank N. Featherstone for proposing the original idea that has lead to this project. 
We wish also to thank A. Palacios and L. Amard for providing the 1D stellar evolution and structure models used as input to setup our 3D ASH simulations.
Special thanks to K. Augustson for providing some of the analysis routines used in section \ref{MFG}.
We acknowledge A. Strugarek for the 3D visualization of the 40\% model. 
We also thank V. Réville for the analysis of the evolution of the Alfvén radius and loss of angular momentum in our MHD simulations. 
We acknowledge funding support by ERC STARS2 207430, CNES Solar Orbiter, Plato and INSU/PNST fundings.


\bibliographystyle{aasjournal} 
\bibliography{EVB2017_1} 

\end{document}